\documentclass[reprint,aps,pra,nofootinbib,superscriptaddress]{revtex4-2}

\usepackage{braket}
\usepackage{slashed}
\usepackage{graphicx}
\usepackage{xcolor}
\usepackage{enumitem}
\usepackage{siunitx}
\usepackage{mathtools}
\usepackage{dsfont}
\usepackage[hidelinks]{hyperref}
\usepackage{amsmath,amsthm,amsfonts,amscd,amssymb} 
\usepackage{comment}

\usepackage{tikz} 
\usetikzlibrary{shapes,arrows,positioning,automata,backgrounds,calc,er,patterns}
\usetikzlibrary{decorations,decorations.markings,decorations.pathreplacing,decorations.pathmorphing}
\usepackage{relsize}

\newcommand{\be}{\begin{equation}}
\newcommand{\ee}{\end{equation}} 
\newcommand{\mb}{\mathbf}

\newcommand{\bes}[1]{\begin{equation} \begin{split} #1 \end{split} \end{equation}}

\DeclareMathOperator{\tr}{tr}

\begin{document}

\title{On the quantum mechanics of entropic forces}

\author{Daniel Carney}
\affiliation{Physics Division, Lawrence Berkeley National Laboratory, Berkeley, CA 94720}
\author{Manthos Karydas}
\affiliation{Physics Division, Lawrence Berkeley National Laboratory, Berkeley, CA 94720}
\author{Thilo Scharnhorst}
\affiliation{Physics Division, Lawrence Berkeley National Laboratory, Berkeley, CA 94720}
\affiliation{Department of Physics, University of California, Berkeley, CA 94720}
\author{Roshni Singh}
\affiliation{Physics Division, Lawrence Berkeley National Laboratory, Berkeley, CA 94720}
\affiliation{Department of Physics, University of California, Berkeley, CA 94720}
\author{Jacob M. Taylor}
\affiliation{Joint Center for Quantum Information and Computer Science, University of Maryland, College Park, MD 20742}
\affiliation{Joint Quantum Institute, National Institute of Standards and Technology, Gaithersburg, MD 20899}

\begin{abstract}

It was conjectured thirty years ago that gravity could arise from the entropic re-arrangement of information. In this paper, we offer a set of  microscopic quantum models which realize this idea in detail. In particular, we suggest a simple mechanism by which Newton's law of gravity arises from extremization of the free energy of a collection of qubits or oscillators, rather than from the exchange of virtual quanta of a fundamental field. We give both a local and non-local version of the construction, and show how to distinguish a range of these entropic models from ordinary perturbative quantum gravity using existing observations and near-term experiments.

\end{abstract}

\maketitle

\tableofcontents

\section{Introduction}

While many forces in nature are mediated by exchange of virtual field quanta, there are also effective forces which arise from complex systems driving thermodynamic free energies to their extrema. In particular, it has been suggested that gravity could arise as such a thermal or entropic interaction, rather than through a fundamental quantum field~\cite{jacobson1995thermodynamics,verlinde2011origin}. In this scenario, it has been unclear how quantized matter would couple to gravity. Here, we study this issue by constructing detailed, fully quantum-mechanical models which reproduce Newton's law of gravitation in the thermodynamic limit.

\begin{figure}[t]

\begin{tikzpicture}

\draw (0,-1) -- (0,1) -- (1,2) -- (1,0) -- (0,-1);
\draw (3,-1) -- (3,1) -- (4,2) -- (4,0) -- (3,-1);
\node at (3.7,.5) {$A$};
\draw [thick,<->] (.7,1.5) -- (3.3,1.5);
\node at (2,1.8) {$x$};

\draw [fill=lightgray] (.6,.2) circle (.1);
\draw [fill=lightgray] (.9,.8) circle (.1);
\draw [fill=lightgray] (1.2,.5) circle (.1);
\draw [fill=lightgray] (1.4,-.2) circle (.1);
\draw [fill=lightgray] (1.9,.6) circle (.1);
\draw [fill=lightgray] (2.1,-.1) circle (.1);
\draw [fill=lightgray] (2.4,.2) circle (.1);
\draw [fill=lightgray] (2.5,-.3) circle (.1);
\draw [fill=lightgray] (2.9,.8) circle (.1);
\draw [fill=lightgray,opacity=.5] (3.2,.4) circle (.1);
\draw [fill=lightgray,opacity=.5] (3.4,-.1) circle (.1);

\draw [<->,dashed] (1,-.8) -- (1,-1.8);
\draw [<->,dashed] (1.75,-.8) -- (1.75,-1.8);
\draw [<->,dashed] (2.5,-.8) -- (2.5,-1.8);

\draw (.8,-2) rectangle (2.7,-3);
\node at (1.75,-2.5) {bath};

\end{tikzpicture}

\caption{The ideal gas law as an entropic force between two movable walls.}
\label{fig:gas}
\end{figure}
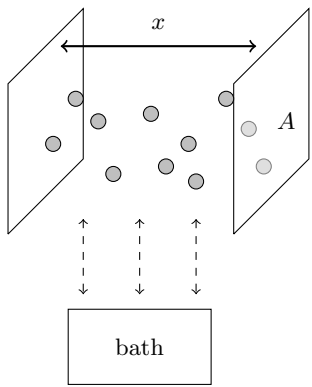

The essence of the idea can be understood by analogy with a typical entropic force, like the ideal gas law. Consider a pair of massive pistons with a non-interacting gas between them, as in Fig.~\ref{fig:gas}. We assume the gas is held at fixed temperature $T$ by connection to a heat bath. The free energy of the gas $\mathcal{A} = U - TS$ depends on the distance between the pistons $x$ by the usual Sackur-Tetrode formula,
\be
\frac{\partial S}{\partial x} = \frac{N}{x}.
\ee
The system is entropically driven towards the extrema of its free energy, which is determined by $\partial \mathcal{A}/\partial x = 0$. Since $\partial U/\partial x=0$ in the ideal gas, this means that the piston separation is driven by the entropic force
\be
F = T \frac{\partial S}{\partial x} = \frac{N T}{x} = P A
\ee
where $A$ is the area of the pistons. What this exercise demonstrates is that the two pistons feel an effective force between them, namely the pressure, which is mediated by the gas rather than some fundamental quantized field.

In our models, the two-body Newtonian gravitational interaction $\mb{F} = -G_N m_1 m_2 \hat{\mb{r}}/r^2$ will similarly arise as a pressure mediated by a microscopic system which is driven toward extremization of its free energy. Our primary goal is to demonstrate how non-relativistic gravity can arise in detail as a thermodynamic limit of a controlled microscopic model. This in turn can explain how gravity couples to quantized matter in such a scenario (a question left open by the treatments in~\cite{jacobson1995thermodynamics,verlinde2011origin}) and how such a scenario can be experimentally distinguished from ordinary virtual graviton exchange~\cite{Carney:2018ofe,Lindner:2004bw,Kafri:2013wxa,Kafri:2014zsa,bose2017spin,marletto2017gravitationally,Matsumura:2020law,Datta:2021ywm,Carney:2021yfw,Oppenheim:2022xjr,Lami:2023gmz}. We discuss the experimental predictions in Sec. \ref{sec:experiments}.

We emphasize that our construction does not necessarily encompass all possible realizations of the proposal that gravity arises as an entropic or thermal effect~\cite{jacobson1995thermodynamics,verlinde2011origin}. In particular, our models generally have both thermal and purely entropic components to the force, consistent with \cite{jacobson1995thermodynamics} but more general than \cite{verlinde2011origin}. We also make a number of specific choices (e.g., bath spectral densities) in order to give detailed calculations, but we expect that the qualitative structure we find here will hold in a much broader setting.

The mechanism presented here is also different from the way that gravity emerges in holographic models like AdS/CFT~\cite{Maldacena:1997re,Lashkari:2013koa}, at least at face value. Here gravity arises as a direct interaction mediated by a thermal system, whereas any string theoretic model of AdS/CFT has massless spin-2 quanta in the low energy limit, i.e., gravitons, which act as a coherent quantum mediator of gravity at long wavelength~\cite{weinberg1964photons,weinberg1965photons,Carney:2021vvt}.

To summarize, our goal is to provide explicit, proof-of-concept examples in which gravity emerges as an entropic effect, in a \emph{different} fashion than the usual graviton picture, and thus also differently from standard holography. Interestingly, however, we find that the models have a range of free parameters, and in some parameter regimes become indistinguishable from standard virtual graviton exchange. This may be an artifact of our non-relativistic limit, or it may be more fundamental. We leave this important issue to future work.

Before moving on, we note some helpful earlier work on entropic gravity~\cite{hossenfelder2010comments,Carroll:2016lku,schimmoller2021decoherence} and on quantum treatments of entropic forces~\cite{sokolov2010statistical,wang2016quantum}. Throughout the paper we use units $k_{\rm B} = \hbar = c = 1$.

\section{Newton's law of gravitation}

In the ideal gas example, the force on the pistons arises through a coupling to a many-body mediator system (the gas molecules), which is in turn coupled to a heat bath. We will follow the same paradigm in order to construct a model of gravity. The Hamiltonian will take the general form
\be
\label{eq:H-full}
H = H_S + H_M + V_{SM} + H_B + V_{MB}.
\ee
Here, $S$ refers to the observable system of non-relativistic massive bodies, with
\be
H_S = \sum_{i=1}^{N} \frac{\mb{p}_i^2}{2m_i}.
\ee
An external potential can be added trivially. The label $M$ refers to a many-body mediator system to be discussed in detail below, and $B$ refers to the heat bath. Note that the massive bodies do not couple directly to the bath. 

In the ideal gas example, the bath would represent a thermal reservoir, for example thermal phonons in the walls of the chamber, which can be exchanged with the gas. The role of the bath is to fix the mediator state to temperature $T$, at least on sufficiently long time scales. We make this precise below. The introduction of a bath is mostly for calculational convenience: one could also just take the mediators to be a self-thermalizing system, like a self-interacting gas~\cite{ODonovan:2024ocu}.

Newton's law of gravity will arise in the thermodynamic limit. One of our central goals in this paper is to show how this can be understood in the fully quantum setting, including the quantum mechanics of the massive bodies. However, to begin, consider the limit that the masses are heavy enough that we can view them as classical objects, so the positions $\mb{x}_i$ are c-numbers. Here is the essential idea. The free energy $\mathcal{A} = U - TS$ of the mediator will depend on the mass positions $\mathcal{A} = \mathcal{A}(\mb{x}_1, \mb{x}_2, \ldots)$ through the coupling $V_{SM}$. Then the $i$th mass feels a thermal force
\be
\label{eq:F-general}
\mb{F}_i = - \nabla_{\mb{x}_i} \mathcal{A}
\ee
exerted by the mediator as it tries to extremize its free energy. What we will do in the next two sections is give examples of the mediator $H_M$ and its coupling to the masses $V_{SM}$ such that
\be
\label{eq:Newton-classical}
-\nabla_{\mb{x}_i} \mathcal{A} = -\sum_{j \neq i} G_N m_i m_j \frac{\hat{\mb{x}}_{ij}}{|\mb{x}_{ij}|^2},
\ee
where $\mb{x}_{ij} = \mb{x}_i - \mb{x}_j$. Newton's constant $G_N$ will emerge from some specific combinations of the microscopic parameters defining the mediator and its coupling.

Quantum mechanically, the force on the $i$th body due to the mediator is given by the Heisenberg equation
\be
\mb{F}_i = \dot{\mb{p}}_i = i [V_{SM}, \mb{p}_i].
\ee
Thus, the quantum version of the classical force law in Eq.~\eqref{eq:Newton-classical} will require us to find the mediator and coupling such that
\be
\label{eq:Newton-quantum}
\tr_{MB} \left( \rho_{MB} \mb{F}_i \right) = -\sum_{j \neq i} G_N m_i m_j \frac{\hat{\mb{x}}_{ij}}{|\mb{x}_{ij}|^2} + \cdots,
\ee
where now $\mb{x}_i$ and $\mb{p}_i$ are operators, and $\rho_{MB} = \tr_S \rho$ is the density matrix for the mediator and bath. Here the dots represent corrections due to thermal fluctuations, as we will see below. This equation will hold when the mediator is in a thermal or nearly thermal state at temperature $T$, but our microscopic construction also contains a vast set of far-from-equilibrium states. Our goal is to realize an idea like that of~\cite{jacobson1995thermodynamics}, where semiclassical gravity arises from adiabatic evolution between different equilibrium states of the mediator system.

\subsection{Non-local model}
\label{sec:general-construction}

\begin{figure}[t]

\begin{tikzpicture}

\draw [dashed,thick,decorate,decoration={snake,segment length=1.3cm}] (-3,0)  [out=45,in=135] to (3,0);
\draw [thick,decorate,decoration={snake,segment length=1.1cm}] (-3,0)  [out=35,in=145] to (3,0);
\draw [dashed,thick,decorate,decoration={snake,segment length=0.9cm}] (-3,0)  [out=25,in=155] to (3,0);
\draw [dashed,thick,decorate,decoration={snake,segment length=0.7cm}] (-3,0)  [out=15,in=165] to (3,0);
\node at (0,-.1) {$\vdots$};
\draw [dashed,thick,decorate,decoration={snake,segment length=0.5cm}] (-3,0)  [out=-25,in=205] to (3,0);
\draw [thick,decorate,decoration={snake,segment length=0.3cm}] (-3,0)  [out=-45,in=225] to (3,0);

\draw [fill=lightgray] (-3,0) circle (1);
\draw [fill=lightgray] (3,0) circle (1);

\node at (-3,0) {$m_1$};
\node at (3,0) {$m_2$};

\end{tikzpicture}

\caption{\textbf{Non-local spin entropic gravity.} One possible visualization of the non-local entropic gravitational interaction. Each line represents one qubit $\alpha$, with the frequency $\omega_{\alpha}$ depicted by the wavelength. The qubits are thermally occupied, with $\ket{0}_{\alpha}$ shown as a dashed line and $\ket{1}_{\alpha}$ as a solid line. The heat bath is not shown but could be included analogously to that in Fig.~\ref{fig:gas}.}
\label{fig:cartoon}
\end{figure}
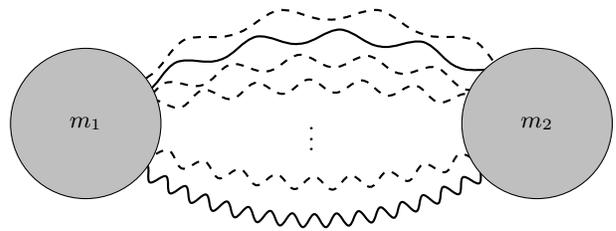

First we present a non-local model. Consider two massive bodies with masses $m_1, m_2$ and relative position $\mb{x} = \mb{x}_1 - \mb{x}_2$. We generalize to an arbitrary number of bodies shortly. For the mediating system, we introduce a discrete set of qubits labeled by $\alpha=1,2,3,\dots$. Their  frequencies depend on the relative position,
\be
\label{eq:H-first}
V_{SM} = \sum_{\alpha=1}^{\infty} \omega_{\alpha}(\mb{x}) N_{\alpha}, \ \ \ N_{\alpha} = \frac{1-Z_{\alpha}}{2}, 
\ee
where $Z_{\alpha}$ is a Pauli matrix. We have chosen conventions so that $\ket{0}$ is the lowest-energy state and has zero energy. The mediator frequencies are taken to have a linear spectrum:
\be
\label{eq:omega-f}
\omega_{\alpha}(\mb{x}) = \alpha\, f(\mb{x}).
\ee
At this stage, $f(\mb{x})$ is an arbitrary function of $\mb{x}$ with units of energy. For intuition, we give a possible graphical representation in Fig.~\ref{fig:cartoon}. This coupling is non-local in the sense that the mediator qubits all couple directly to $\mb{x} = \mb{x}_1 - \mb{x}_2$, rather than separately to $\mb{x}_1$ and $\mb{x}_2$.

The key idea is that the function $f(\mb{x})$ can be chosen appropriately so that the Newtonian force, or any other conservative central force, arises in the thermodynamic limit. The same construction also works equally well with the qubits replaced by oscillators, and $N_{\alpha} = a_{\alpha}^\dag a_{\alpha}$. In the rest of the paper we specialize to the qubits for computational convenience. To emphasize the generality of the idea, we also show the results for the oscillator version in Appendix~\ref{appendix:oscillators}. We will refer to the specific constructions with qubit mediators as ``spin entropic gravity'' models.

To see how Newton's law arises, consider the limit where the masses can be treated classically so that $\mb{x}$ is a c-number. Suppose the qubits are prepared in the thermal state,
\be
\label{eq:rhoT}
\rho_{M} = \bigotimes_{\alpha=1}^{\infty} \rho_{T,\alpha,\mb{x}}, \ \ \rho_{T,\alpha,\mb{x}} = Z^{-1}_{T,\alpha,\mb{x}} \begin{pmatrix} 1 & 0 \\ 0 & e^{-\omega_{\alpha}(\mb{x})/T} \end{pmatrix},
\ee
where the partition function of each qubit is $Z_{T,\alpha,\mb{x}} = 1 + e^{-\omega_{\alpha}(\mb{x})/T}$. The free energy and partition function of the full set of qubits are thus
\begin{align}
\label{eq:Partition_function_Free_energy_local}
\mathcal{A} = - T \ln Z, \ \ \ Z  = \prod_{\alpha} \left[ 1 + e^{-\omega_{\alpha}(\mb{x})/T} \right].
\end{align}
Now we take the usual thermodynamic limit, by assuming that we have many qubits and can take a continuum limit $\sum_{\alpha} \to \int d\alpha= \int d\omega/f(\mb{x})$. This gives
\begin{align}
\label{eq:FE_local}
\mathcal{A} = - T \int_{0}^{\infty} \frac{d\omega}{f(\textbf{x})} \ln ( 1 + e^{-\omega/T}) = -\frac{\pi^2}{12} \frac{ T^2}{ f(\textbf{x})}.
\end{align}
The masses therefore feel an effective thermal force, analogous to the gas pressure on the walls discussed above, given by
\be
\label{eq:FN}
\mb{F}_{\rm th} = -\nabla \mathcal{A} =\frac{\pi^2}{12} T^2 \nabla\left(\frac{1}{f(\mb{x})}\right).
\ee
Thus, if we choose the mediator frequencies to have dependence on the mass positions in the form
\be
\label{eq:f(x)}
\frac{1}{f(\mb{x})}= \lambda + \frac{\ell^{2}}{|\mb{x}|}\,,
\ee
where $\lambda,\ell$ are some constants with dimensions of length, we obtain
\be
\label{eq:Fth}
\mb{F}_{\rm th} =- \frac{\pi^{2}}{12} T^{2}\ell^{2} \frac{\hat{\mb{x}}}{|\mb{x}|^2}= -G_N m_1 m_2 \frac{\hat{\mb{x}}}{|\mb{x}|^2},
\ee
where, to match to Newton's law, we take
\be
\label{eq:T-G}
\frac{\pi^2}{12} T^2 \ell^2 \equiv G_N m_1 m_2.
\ee
With this identification, we obtain Newton's law of gravity from the interaction with a collection of thermally-driven qubits.

The ontological status of the identification in Eq.~\eqref{eq:T-G} is somewhat uncertain. We will say that $G_N$ ``emerges'' from the microscopic parameters $T,\ell$. At a basic level we recognize that this identification is dimensionally consistent. It should also be emphasized that this fixes the product $T \ell$ of two free parameters in our model in terms of observable gravitational quantities ($G_N$ and the masses). The other free parameter, $\lambda$, does not appear in Eq. \eqref{eq:Fth} and thus its existence is not strictly necessary in order to recover Newton's law. We will see how these parameter degeneracies are broken later when we study noise in the model.

In the introduction, we mentioned that this force has both thermal and entropic components. To be more explicit, from Eq.~\eqref{eq:Partition_function_Free_energy_local} one finds the internal energy and entropy of the mediators:
\be
\label{eq:U-total}
U = \frac{\pi^2}{12} \frac{T^2}{f(\mb{x})}, \ \ \ S = \frac{\pi^2 }{6} \frac{T}{f(\mb{x})}.
\ee
Thus we see that the free energy gradient $\nabla \mathcal{A}$ contains both thermal and entropic components. This is unlike the ideal gas, which has $\nabla U = 0$. 

Our result for the force, \eqref{eq:Fth}, holds assuming that the masses are can be treated classically, and assuming that the qubits are held in thermal equilibrium. To view \eqref{eq:Fth} as a dynamical law, we have to assume that the mediator qubits remain thermalized as the distance between the bodies changes. This requires that the motion of the massive bodies is in some sense adiabatic, meaning that it is slow compared to some thermalization timescale in the gas. This is similar to the way an ideal gas is assumed to remain in the thermal state as piston walls are moved, and is the same approximation considered when deriving the Einstein equation as an equation of state in Ref.~\cite{jacobson1995thermodynamics}. In Sec.~\ref{sec:bath-lindblad}, we show how to implement this with a bath-mediator coupling.

This discussion leaves open the question of how to extend this treatment to the fully quantum situation, where the positions $\mb{x}_{1,2}$ of the masses become operators. Using the full Hamiltonian of Eq. \eqref{eq:H-full}, we have the Heisenberg equations of motion for the masses:
\be
\label{eq:xp-EOM}
\dot{\mb{x}}_{i} = \frac{\mb{p}_{i}}{m_{i}}, \ \ \ \dot{\mb{p}}_{i} = - \sum_{\alpha} \nabla_{\mb{x}_i} \omega_{\alpha}(\mb{x}) N_{\alpha}.
\ee
Consider a product state of the form $\rho \approx \ket{\psi} \bra{\psi} \otimes \rho_{M}$, where $\ket{\psi}$ is a localized around some position $\mb{x}$, and $\rho_{M}$ is the thermal state of the qubits with frequencies set by this value of $\mb{x}$ given in Eq. \eqref{eq:rhoT}. Using exactly the same kind of manipulations as in the thermodynamics case above, we have the equation of motion for the relative momentum $\mb{p}$:
\begin{align}
\label{eq:FN-th}
\braket{ \dot{\mb{p}} }  = - \sum_{\alpha} \Braket{ \nabla_{\mb{x}} \omega_\alpha(\mb{x}) N_{\alpha} } = -G_N m_1 m_2 \Braket{ \frac{\hat{\mb{x}}}{ \mb{x}^2} }.
\end{align}
The point here is that we can see the same thermodynamic force showing up in the quantum Heisenberg equations, assuming the masses are well-localized and  the mediator is averaged over. 

Finally, to generalize to $N$ masses, there are a number of possible methods. The simplest is to just repeat the construction for every pair $i,j$. Concretely, we replace the Hamiltonian of Eq.~\eqref{eq:H-first} with
\be
V_{SM} = \sum_{i \neq j} \sum_{\alpha=1}^{\infty} \omega_{ij,\alpha}(\mb{x}_{ij}) N_{ij,\alpha},
\ee
where $\mb{x}_{ij} = \mb{x}_i - \mb{x}_j$ and now each pair has its own system of mediator qubits with number $N_{ij,\alpha} = (1-Z_{ij,\alpha})/2$. The frequencies are now functions of the relative coordinates:
\be
\omega_{ij,\alpha} = \alpha f_{ij}(\mb{x}_{ij}), \ \ f_{ij} = \lambda + \frac{\ell_i \ell_j}{|\mb{x}_{ij}|},
\ee
with $\mb{x}_{ij} = \mb{x}_i - \mb{x}_j$. To match to the Newtonian force, we introduce both a temperature $T$ and some overall length scale $L$, and identify
\be
\label{eq:T-G-many}
\frac{\pi^2}{12} T^2 \equiv \frac{G_N}{L^4}, \ \ \ \ell_i \equiv m_i L^2,
\ee
which generalizes the two-body identification in Eq.~\eqref{eq:T-G}. There are still only two free parameters, $T$ and $L$, constrained by one equation.

\subsection{Local model}
\label{sec:local-construction}

In the previous section, we showed how to generate an emergent Newtonian force by directly coupling pairs of massive objects to mediator qubits. Here we show an alternative construction, in which each massive object couples quasi-locally to a lattice of background qubits. 

Although some elements of the non-local model from Sec. \ref{sec:general-construction} and the local model shown here are common to each other, they are not two versions of the same model. In particular, unlike integrating photons out to get a non-local Coulomb force, one can not integrate out the local mediator here in order to recover an effective description of the form in Sec. \ref{sec:general-construction}. The two models have some qualitative differences and are subject to different experimental constraints, as we discuss below.

\begin{figure}[t]

\begin{tikzpicture}[scale=.7]

\draw (2,1) -- (2,9);
\draw (4,1) -- (4,9);
\draw (6,1) -- (6,9);
\draw (8,1) -- (8,9);

\draw (1,2) -- (9,2);
\draw (1,4) -- (9,4);
\draw (1,6) -- (9,6);
\draw (1,8) -- (9,8);

\draw [thick,<->] (2,0.5) -- (4,0.5);
\node at (3,0) {$a$};

\draw [fill=lightgray] (6,4) circle (1);
\node at (6,4) {$m$};

\node [rotate=65,scale=2] at (2,2) {$\uparrow$};
\node [rotate=20,scale=2] at (4,2) {$\uparrow$};
\node [rotate=20,scale=2] at (6,2) {$\uparrow$};
\node [rotate=20,scale=2] at (8,2) {$\uparrow$};

\node [rotate=165,scale=2] at (2,4) {$\uparrow$};
\node [rotate=20,scale=2] at (4,4) {$\uparrow$};
\node [rotate=20,scale=2] at (8,4) {$\uparrow$};

\node [rotate=-75,scale=2] at (2,6) {$\uparrow$};
\node [rotate=20,scale=2] at (4,6) {$\uparrow$};
\node [rotate=20,scale=2] at (6,6) {$\uparrow$};
\node [rotate=20,scale=2] at (8,6) {$\uparrow$};

\node [rotate=220,scale=2] at (2,8) {$\uparrow$};
\node [rotate=60,scale=2] at (4,8) {$\uparrow$};
\node [rotate=310,scale=2] at (6,8) {$\uparrow$};
\node [rotate=140,scale=2] at (8,8) {$\uparrow$};

\end{tikzpicture}

\caption{\textbf{Local spin entropic gravity.} In the local model, space is filled with a lattice of spacing $a$, with one qubit on each site. A mass causes the nearby qubits to slightly polarize with respect to the background, leading to a decrease in the local entropy.}
\label{fig:local-cartoon}
\end{figure}
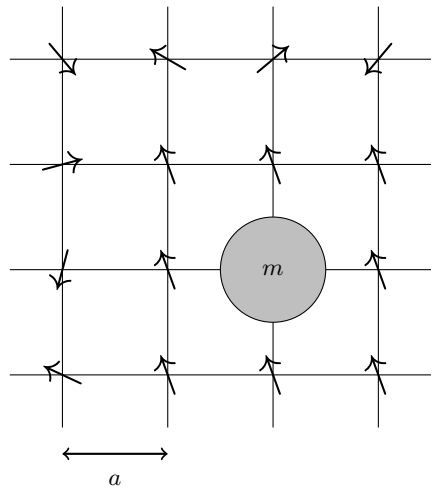

Consider a 3D lattice of spacing $a$. At each site $\alpha$ of the lattice, we have a qubit fixed at the site, with number operator $N_{\alpha}$ as before. See Fig.~\ref{fig:local-cartoon}. We take the Hamiltonian for the qubits
\begin{equation}
V_{SM} = \sum_{\alpha} \Omega_{\alpha}(\mb{x}_1, \mb{x}_2, \ldots) N_{\alpha}.
\end{equation}
As in the non-local model, we encode the positions of the massive bodies in the frequencies of the qubits:
\begin{align}
\label{eq:Omega}
\Omega_{\alpha} &= -\mu + \omega_{\alpha}, \ \ \omega_{\alpha} = \sum_i \frac{\ell_i}{|\mb{r}_{\alpha} - \mb{x}_i|^2 + a^2}.
\end{align}
Here $\mb{r}_{\alpha}$ are the lattice coordinates and $\mb{x}_i$ are the position operators for the massive bodies. The constants $\ell_i = m_i L^2$ as in Eq.~\eqref{eq:T-G-many}, where $L$ is again some overall length scale introduced as a free parameter. The mediator spins near a mass have a larger energy splitting than those farther from a mass. The variable $\mu$ is a chemical potential for the lattice spins. Thus in total we have three free parameters: the length scale $L$, the lattice spacing $a$, and the chemical potential $\mu$.

Consider the Helmholtz free energy of the mediators in equilibrium at temperature $T$. We regulate the calculations by placing the lattice in a total volume $V$, with $V \to \infty$ at the end of the calculation. The partition function and free energy are
\begin{align}
\mathcal{A} = -T \ln Z, \ \ Z = \prod_{\alpha} \left[ 1 + e^{(\mu - \omega_{\alpha})/T} \right].
\end{align}
Now let us consider expanding this in the limit of small frequencies  $\omega_{\alpha} \ll \mu$. The leading terms are
\begin{align}
\label{eq:A-local-taylor}
\mathcal{A} = -\frac{V T}{a^3} \ln Z_* + \sum_{\alpha} \sigma_* \omega_{\alpha} - \frac{1}{2} \sigma_*(1-\sigma_*) \frac{\omega_{\alpha}^2}{T} + \ldots.
\end{align}
Here $Z_* = 1 + e^{\mu/T}$ is the single-site partition function in the absence of any masses, and 
\be
\label{eq:sigmastar}
\sigma_* = \braket{N_{\alpha}} = \frac{1}{e^{-\mu/T} + 1}
\ee
is the average polarization.\footnote{We remark that the $a \rightarrow 0$ limit is well defined in what comes due to the polarization of spins going to zero faster than $\Omega$ goes to infinity. We use finite $a$ for simplicity.}

Let us study the effects of the terms in the free energy. The first term in Eq.~\eqref{eq:A-local-taylor} is a constant function of the $\mb{x}_i$ and does not contribute to any forces on the massive bodies. The second term is also a constant: we have
\be
\sum_{\alpha} \omega_{\alpha} \approx \sum_i \frac{\ell_i}{a^3} \int \frac{d^3\mb{r}}{|\mb{r}+\mb{x}_i|^2+a^2} = {\rm const.},
\ee
where we took the continuum limit $\sum_{\alpha} \to \int d^3\mb{r}/a^3$. This constant can be absorbed into a renormalization of $\mu$, which we implicitly assume in what follows. The Newtonian gravitational force arises from the final $\omega_{\alpha}^2$ term. Using Eq.~\eqref{eq:Omega} in this term, we obtain 
\begin{align}
\begin{split}
\label{eq:Newton-local}
\mathcal{A} & = {\rm const.} + \frac{1}{2} \frac{\sigma_*(1-\sigma_*)}{T} \sum_{\alpha} \omega_{\alpha}^2 \\
& \approx {\rm const.} + \sum_{ij} \frac{\ell_i \ell_j}{2 a^3} \frac{\sigma_*(1-\sigma_*)}{T} \mathcal{I}(\mb{x}_i,\mb{x}_j),
\end{split}
\end{align}
up to terms of $\mathcal{O}(\omega_{\alpha}^3)$. In the second line, we again converted the lattice sum to an integral,
\begin{align}
\label{eq:Newton-local-2}
\mathcal{I} = \int \frac{d^3\mb{r}}{(|\mb{r} - \mb{x}_i|^2 + a^2)(|\mb{r} - \mb{x}_j|^2 + a^2)} = \frac{\pi^3 s(|\mb{x}_{ij}|)}{|\mb{x}_{ij}|}
\end{align}
where
\be
\label{eq:f_local_dfn}
s(x) = 1 - \frac{2}{\pi} \arctan(2 a/x)  \approx  1 - \frac{4 a}{\pi x} + \ldots. 
\ee
To get the explicit form of $s$, one can perform the integral by going to cylindrical coordinates along the axis defined by the relative position $\mb{x}_{ij} = \mb{x}_i - \mb{x}_j$. Note that $s(x)$ softens the $1/r^2$ interaction at short range. In the limit $x \rightarrow 0$, such as when $i = j$, the integral $\mathcal{I} \rightarrow \pi^2/a$.

The results in Eqs.~\eqref{eq:Newton-local}, \eqref{eq:Newton-local-2}, and \eqref{eq:f_local_dfn} show that at distances longer than the lattice spacing $a \ll |\mb{x}_{ij}|$, variations in the free energy lead to an attractive $1/r^2$ force law between the massive objects. We then see the necessary identification of microscopic parameters to obtain Newton's law: 
\begin{equation}
\label{eq:G-sigma}
\frac{\sigma_* (1 - \sigma_*)}{T} \frac{ \pi^3 L^4}{a^3}  \equiv G_N.
\end{equation}
We have lost a factor of two due to the double counting of masses in Eq.~\eqref{eq:Newton-local}.

The basic mechanism in this model is that a mass slightly polarizes a cloud of spins around it, which moves with the mass, as shown in Fig.~\ref{fig:local-cartoon}. This polarization both increases the energy and decreases the entropy $S = -\partial_T \mathcal{A}$ in the region near the mass. As two masses approach each other, there can be an overall increase in entropy as the overall polarized volume decreases when the masses get close, and the polarization starts to saturate near the opposite mass. This increase in entropy leads to an attractive interaction, controlled by the negative sign in front of the $\sum_{\alpha} \omega_{\alpha}^2$ term in the free energy.

Unlike the non-local model, there is a regime in which this local model is a purely entropic force. In the non-local model, we found that for any values of the free parameters, both the entropy and internal energy always contribute to the free energy, as in Eq.~\eqref{eq:U-total}. In the local model, in contrast, one can compare the $\sum_{\alpha} \omega_{\alpha}^2$ terms,
\begin{align}
\begin{split}
U &\approx  \frac{\sigma_* (1-\sigma_*)}{2T} \left[2 + \frac{\mu}{T} (1 - 2 \sigma_*)\right] \left(\sum_{\alpha} \omega_{\alpha}^2\right) \\
S & \approx \frac{\sigma_* (1-\sigma_*)}{2 T^2} \left[1 + \frac{\mu}{T} (1 - 2 \sigma_*)\right] \left(\sum_{\alpha} \omega_{\alpha}^2\right).
\end{split}
\end{align}
We see that there is a specific point where the $\omega^2$ term in the mean energy is zero but the same term in the entropy is non-zero, when $\mu = - T \times 2.39936...$. At this entropic point we have an emergent Newtonian gravity theory in which the central force is purely entropic. As far as we can tell, there is nothing special about this point in the parameter space.

The discussion here was, again, in the limit where the masses can be treated classically. Extending to the fully quantum case follows the same steps as the non-local model of Sec.~\ref{sec:general-construction}. In particular, we still need to couple the mediator qubits to a bath in some way so that the mediator state remains thermalized at temperature $T$ on sufficiently slow time scales. We show how to do this in the next two sections.

\subsection{Full microscopic dynamics}
\label{sec:bath-lindblad}

In the previous sections, we showed how to find a mediator system $M$ and coupling $V_{SM}$ of the mediators to the masses so that, assuming the mediators are in thermal equilibrium, they generate an effective Newtonian gravitational force on the masses. What remains to do is to specify dynamics that also keep the mediators thermalized as the masses move.

Following our discussion above, we now introduce a heat bath $B$ and give it a coupling $V_{MB}$ to the mediators. The key idea will be to introduce a separation of three time scales, schematically $\Gamma_{B} \gg \Gamma_M \gg \ddot{\mb{x}}_i$. Here $\Gamma_B$ is the rate at which the bath thermalizes itself, $\Gamma_M$ is the rate at which the bath thermalizes the mediators, and $\ddot{\mb{x}}_i$ are the accelerations of the massive bodies. This separation of time scales allows us to first integrate out the bath, and then integrate out the mediators, to obtain the reduced equations of motion for just the massive objects. 

We take the bath to consist of a large set of free harmonic oscillators, and take the bath terms appearing in Eq.~\eqref{eq:H-full} to be
\begin{align}
\begin{split}
H_B & = \sum_{\alpha,k} f_{\alpha,k} b_{\alpha,k}^\dag b_{\alpha,k} \\
V_{MB} & =  \sum_{\alpha,k} g_{\alpha,k} \sigma_{\alpha} b_{\alpha,k}^\dag + g_{\alpha,k}^*  \sigma^\dag_{\alpha} b_{\alpha,k}.
\end{split}
\label{eq:bath-terms}
\end{align}
Here $\sigma_{\alpha} = \sigma_{\alpha,-} = \ket{0} \bra{1}_{\alpha}$ is the lowering operator for the $\alpha$th qubit. The bath oscillators couple to the mediators through an exchange coupling, where a mediator qubit can lose one quantum of energy by giving it to the bath, and vice versa. 

Following standard calculations, we can first trace out the bath in the limit that $\Gamma_B$ is the fastest time scale. Invoking the Born-Markov approximation gives an equation of motion for the state $\rho_{SM}$ of the masses and mediator in Lindblad form~\cite{wang2016quantum,Nathan:2020qck},
\begin{align}
\begin{split}
\label{eq:lindblad}
& \dot{\rho}_{SM} = -i [H_S + H_M + V_{SM},\rho_{SM}] \\
& + \sum_{\alpha,\pm} L_{\alpha,\pm} \rho_{SM} L_{\alpha,\pm}^\dag -  \frac{1}{2} \sum_{\alpha,\pm} \left\{ L_{\alpha,\pm}^\dag L_{\alpha,\pm}, \rho_{SM} \right\}.
\end{split}
\end{align}
The form of the Lindblad (``jump'') operators appearing here is somewhat different in the local and non-local models. Here we quote the results, and review the detailed derivations in Appendix \ref{appendix:lindblad}.

In the non-local model of Sec.~\ref{sec:general-construction}, the Lindblad operators are
\begin{align}
\begin{split}
\label{eq:lindblad-ops-nl}
L_{\alpha,+} & = \sqrt{\gamma(\omega_{\alpha}(\mb{x}))\, n_B(\omega_{\alpha}(\mb{x}))} \sigma_{\alpha} \\
L_{\alpha,-} & = \sqrt{\gamma(\omega_{\alpha}(\mb{x}))\, [n_B(\omega_{\alpha}(\mb{x}))+1]} \sigma_{\alpha}^\dag.
\end{split}
\end{align}
These represent events where the mediator qubits either gain or lose energy to the bath, respectively. Here and after, $n_B(\omega) = (e^{\omega/T}-1)^{-1}$ is the Boltzmann distribution at frequency $\omega$. The frequency dependent rate $\gamma(\omega_{\alpha})$ is a detailed function of the bath parameters. We will typically use an Ohmic bath, which has damping
\be
\label{eq:ohmic}
\gamma(\omega)=\zeta \omega,
\ee
where $\zeta$ is a dimensionless parameter that controls the damping rate. In Eq.~\eqref{eq:lindblad-ops-nl}, we wrote the operators in terms of one relative coordinate $\mb{x}$; with many bodies $\mb{x}_i$ we add a set of Lindblad operators for each $\mb{x}_{ij}$. Thus the non-local spin entropic gravity model, in total, has four free parameters: $T$, $L$, $\lambda$, and $\zeta$, constrained by one equation [Eq.~\eqref{eq:T-G-many}] which defines the emergent $G_N$.

In the local model of Sec.~\ref{sec:local-construction}, on the other hand, the Lindblad operators take the form
\begin{align}
\begin{split}
L_{\alpha,+} & = \sqrt{\gamma_{\rm th}} e^{-\omega_{\alpha}(\mb{x})/2T} \sigma_{\alpha} \\
L_{\alpha,-} & = \sqrt{\gamma_{\rm th}} \sigma_{\alpha}^\dag.
\end{split}
\end{align}
Here, we have continued to assume that the chemical potential $\mu \gg \omega_{\alpha}$ for the frequencies of interest. Thus the thermalization rate $\gamma_{\rm th}$ is a simple constant, set by the bath density of states at $\omega = \mu$. Here $\mb{x}$ refers to the full set of mass locations $\mb{x}_i$ as in Eq.~\eqref{eq:Omega}. Thus this model also has four free parameters $T$, $\mu$, $a$, and $\gamma_{\rm th}$, constrained by the definition of the emergent $G_N$ in Eq.~\eqref{eq:G-sigma}.

Equation~\eqref{eq:lindblad} is local in time, a result of the Born-Markov approximation. It is the open systems version of a Schr\"odinger equation and in particular preserves the trace $\tr \rho_{SM} = 1$, i.e., conserves total probability. Eq. \eqref{eq:lindblad} gives the detailed quantum evolution of both the massive bodies and mediators on timescales long compared to the thermalization rate $\Gamma_B$ of the heat bath. In particular, Eq. \eqref{eq:lindblad} includes dynamics which drive the qubit mediators out of equilibrium. We study some experimental implications of these fluctuations in Sec. \ref{sec:experiments}.

\subsection{Evolution of the masses in the adiabatic limit}

Finally, we can now integrate out the mediators, and derive an equation of motion for just the state of the massive objects $\rho_M$. To do this, we again invoke our separation of timescales, assuming that the mediator is driven towards equilibrium on a timescale fast compared to the motion of the masses. 

While the heat bath was assumed to reset to the same thermal state at each time step, it is crucial that the mediator state changes in a way contingent on the state of the masses. For example, in the ideal gas, this corresponds to a limit where the walls are moving and thus changing the length $x$, but slowly enough that the gas is evolving through a sequence of equilibrium states plus small perturbations.

With the above assumptions, the equation of motion for the state $\rho_S = \tr_{MB} \rho$ of the massive bodies, on timescales longer than the mediator-bath coupling scale $\Gamma_M$, is again a Lindblad equation. Deriving this equation can be done with a technique known as adiabatic elimination~\cite{carmichael2013statistical,carmichael2007statistical,sakurai2020modern,azouit2016adiabatic}, which essentially amounts to identifying a slowly-decaying mediator mode and averaging over the fast fluctuations around it. Here we give the results, and give a detailed derivation in Appendix~\ref{app:adiabatic}.

In both models, the effective Lindblad equation for the masses takes the form
\begin{align}
\begin{split}
\label{eq:lindblad-final}
\dot{\rho}_S & = -i \left[ H_{\rm eff}, \rho_S \right] \\
& + \sum_{\alpha,\pm} K_{\alpha,\pm} \rho_S K_{\alpha,\pm}^\dag - \frac{1}{2}\sum_{\alpha,\pm} \left\{ K_{\alpha,\pm}^\dag K_{\alpha,\pm}, \rho_S \right\}.
\end{split}
\end{align}
The Hamiltonian term
\be
H_{\rm eff} = \sum_i \frac{\mb{p}_i^2}{2m_i} + V_{N,{\rm ent}}
\ee
includes both the kinetic energy and an ``entropic Newton interaction'', while the other terms represent noise acting on the massive bodies.

In the non-local model, the explicit forms of the operators are as follows. The Newtonian interaction is given by
\begin{align}
\begin{split}
\label{eq:V_N_ent_local_model}
V_{N,{\rm ent}} & = - \sum_{\alpha} n_{FD}[\omega_{\alpha}(\mb{x})] \omega_{\alpha}(\mb{x}) \\
& = - \frac{G_N m_1 m_2}{|\mb{x}|} + {\rm const.}
\end{split}
\end{align}
Here $n_{FD} = n_B/(1+2 n_B)$ is the Fermi-Dirac distribution. As discussed in Appendix~\ref{app:adiabatic}, which reproduces the thermal internal energy in the Newton interaction as in Eq.~\eqref{eq:U-total}.\footnote{We have only been able to precisely derive this in the limit that we are considering massive objects whose individual wavepackets are small compared to the average distances between the objects, i.e., density matrix elements for which the relative positions satisfy $|\mb{x} - \mb{x}'| \ll |\mb{x}|$, which is the situation in realistic experiments. Examining its validity beyond this regime is of clear interest, but we leave this for future work. More details are given in Appendix~\ref{app:adiabatic}.} For the case of an Ohmic bath $\gamma(\omega)=\zeta \omega$, the Lindblad operators can be calculated explicitly. The result is
\begin{align}
\begin{split}
\label{eq:Kpm}
K_{\alpha,+} & = \sqrt{2\zeta T} \int_0^{\omega_{\alpha}(\mb{x})/T} d\nu \, \sqrt{g_{+}(\nu)} \\
K_{\alpha,-} & = \sqrt{\frac{2T}{\zeta}} \int_0^{\omega_{\alpha}(\mb{x})/T} d\nu \, \sqrt{g_{-}(\nu)},
\end{split}
\end{align}
where the dimensionless functions $g_{\pm}$ are given in Eqs.~\eqref{eq:gplus} and \eqref{eq:gminus}. The detailed form of these functions will not be important other than their asymptotic behavior, as we will see below. The crucial point is that one noise operator $\sim \zeta$ while the other $\sim 1/\zeta$, where $\zeta$ controls the damping rate of the mediators. Note also that the noise operators $K_{\alpha,\pm}$ are diagonal in the position basis, so they act to decohere quantum spatial superpositions of the masses into classical probability distributions in position space. 

In the local model, the Newtonian term takes a similar form,
\begin{align}
\begin{split}
\label{eq:VN_local}
V_{N,{\rm ent}} & =  - \sum_{\alpha} \sigma_* \omega_{\alpha}(\mb{x}) + \frac{\sigma_*(1-\sigma_*)}{2T} \omega_{\alpha}^2(\mb{x}) + \cdots \\
& = -\frac{G_N m_1 m_2}{|\mb{x}|} + {\rm const.}
\end{split}
\end{align}
Here the constant $\sigma_*$ [Eq.~\eqref{eq:sigmastar}] plays a similar role as the Fermi-Dirac distribution in Eq.~\eqref{eq:V_N_ent_local_model}, i.e., it reproduces the thermal averages, and again we took the continuum limit to get the second line. Continuing to make our assumption that the bath damping rate in this model is a constant $\gamma_{\rm th}$, the Lindblad operators are quite simple:
\begin{align}
\begin{split}
K_{\alpha,+} & = \sqrt{\frac{\sigma_* \gamma_{\rm th}}{4 T^2} } \omega_{\alpha}(\mb{x}) \\
K_{\alpha,-} & = \sqrt{\frac{2\sigma_* (\sigma_* - 1)^2}{\gamma_{\rm th}}} \omega_{\alpha}(\mb{x}).
\label{eq:lindblad-op-local}
\end{split}
\end{align}
This has a similar structure as the non-local case above: one set of noise operators $\sim \gamma_{\rm th}$ while the other $\sim 1/\gamma_{\rm th}$, and the noise operators are diagonal in the position basis.

Eq.~\eqref{eq:lindblad-final} is the central result of this paper. It gives a time-local evolution law for point masses interacting through an entropic gravitational force. The Hamiltonian term acts as a \emph{coherent} Newtonian interaction: it generates unitary evolution, just like a literal $V =- G_{N}m_1 m_2/|\mb{x}|$ potential operator, and thus just like virtual graviton exchange. However, the Lindbladian terms generate noise, and the masses evolve as an open system. This is very different from standard perturbative quantum gravity, in which the Newton potential operator is the only relevant term, leading to a reversible, unitary evolution law. These two scenarios are observably distinguishable, as we discuss in the next section.

\section{Observable consequences}
\label{sec:experiments}

\begin{figure*}[ht!]

\centering
\begin{tabular}{ccc}

\begin{tikzpicture}[scale=.35]


\draw [dashed] (-5,6) -- (-4.5,2.5);
\draw [black,pattern=north west lines,rotate around={9:(-5,4.5)}] (-5.5,-2.5) rectangle (-4.5,2.5);
\draw [dashed] (-5,6) -- (-5.5,2.5);
\draw [black,pattern=north west lines,rotate around={-3:(-5,4.5)}] (-5.5,-2.5) rectangle (-4.5,2.5);

\draw [black,fill=lightgray] (-5.5,-2.5) rectangle (-4.5,2.5);
\draw (-5,2.5) -- (-5,6);
\draw (-7,8) -- (-3,8);
\draw (-5,6) -- (-6,8);
\draw (-5,6) -- (-4,8);

\node at (-7,2.5) {$m_1$};


\draw [dashed] (5,6) -- (4.5,2.5);
\draw [black,pattern=north west lines,rotate around={-7:(5,4.5)}] (5.5,-2.5) rectangle (4.5,2.5);
\draw [dashed] (5,6) -- (5.5,2.5);
\draw [black,pattern=north west lines,rotate around={4:(5,4.5)}] (5.5,-2.5) rectangle (4.5,2.5);

\draw [black,fill=lightgray] (4.5,-2.5) rectangle (5.5,2.5);
\draw (5,2.5) -- (5,6);
\draw (3,8) -- (7,8);
\draw (5,6) -- (4,8);
\draw (5,6) -- (6,8);

\node at (3,2.5) {$m_2$};

\draw [<->] (-5,-3.5) to (5,-3.5);
\node at (0,-2.5) {$d$};

\end{tikzpicture}
 
 \hspace{.5cm}
 
 &

 \hspace{.5cm}


\begin{tikzpicture}[scale=.75]


\draw [fill=lightgray] (0,0) circle (2);
\node at (-0.5,-0.5) {$m_1$};

\draw [black,pattern=north west lines] (0,2.5) circle (0.1);
\draw [fill=lightgray] (0,3) circle (0.1);
\node at (-0.5,3.5) {$m_2$};

\node at (0.6,2.4) {$\ket{\mb{x}_2}$};
\node at (0.6,3.1) {$\ket{\mb{x}_2'}$};


\draw [thick,->] (0,0) -- (0,2.35);
\node at (-0.5,1.25) {$\mb{d}$};

\draw [thick,->] (-.4,2.5) -- (-.4,3);
\node at (-.8,2.75) {$\delta \mb{x}$};

\end{tikzpicture}

 \hspace{.5cm}

& 

 \hspace{.5cm}


\begin{tikzpicture}[scale=0.35]


\draw [->,thick] (-9,-8) -- (-9,-4);
\node at (-10,-6) {$z$};


\draw [->,dashed] (-3,0) -- (-3,-5);
\draw [dashed] (-3,-5) -- (-3,-10);
\draw [->,dashed] (-7,0) -- (-7,-5);
\draw [dashed] (-7,-5) -- (-7,-10);

\node [scale=.7] at (-5.7,-3.3) {$\ket{L}_1$};
\node [scale=.7] at (-1.7,-3.3) {$\ket{R}_1$};

\draw [black,pattern=north west lines] (-3,-2) circle (.5);
\draw [fill=lightgray] (-7,-2) circle (.5);


\draw [->,dashed] (3,0) -- (3,-5);
\draw [dashed] (3,-5) -- (3,-10);
\draw [->,dashed] (7,0) -- (7,-5);
\draw [dashed] (7,-5) -- (7,-10);

\node [scale=.7] at (8.3,-3.3) {$\ket{R}_2$};
\node [scale=.7] at (4.3,-3.3) {$\ket{L}_2$};

\draw [fill=lightgray] (3,-2) circle (.5);
\draw [black,pattern=north west lines] (7,-2) circle (.5);


\draw [<->] (-5,-8) -- (5,-8);
\node [scale=.8] at (0,-9) {$d$};

\draw [<->] (-3.25,-9) -- (-6.75,-9);
\node [scale=.8] at (-5,-10) {$\delta x$};

\end{tikzpicture}

\end{tabular}

\caption{\textbf{Experiments discussed in Sec.~\ref{sec:experiments}.} Left: two mechanical sensors placed at distance $d$, used to measure their mutual gravitational interaction. Fluctuations in the force are schematically depicted by the dashed masses (Sec.~\ref{sec:noise}). Center: a spatially superposed mass, here depicted near some other source mass, for example an atom interferometer near the surface of Earth (Sec.~\ref{sec:decoherence}). Right: a pair of free-falling masses, both initially superposed into two locations, used to test for the generation of entanglement via gravity (Sec.~\ref{sec:entanglement}).}
\label{fig:experiments}

\end{figure*}
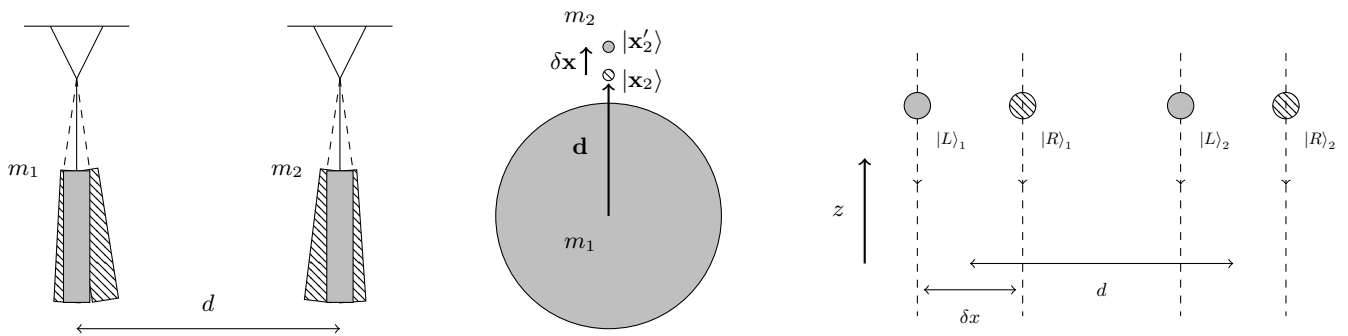

In the past decade, there have been two general classes of experiments proposed to test quantum properties of the gravitational field~\cite{Carney:2018ofe}. The basic idea is to distinguish predictions that are unique to standard perturbative quantum gravity, where low-energy gravitational interactions are mediated by the exchange of gravitons, from a range of proposed alternative models.

Perhaps the best known class of experiments aims to test if the gravitational interaction can generate entanglement between two massive objects~\cite{Lindner:2004bw,Kafri:2013wxa,Kafri:2014zsa,bose2017spin,marletto2017gravitationally,Matsumura:2020law,Datta:2021ywm}. Virtual graviton exchange, or equivalently a coherent Newtonian two-body potential, does entangle masses~\cite{Carney:2018ofe,Carney:2021vvt}. In contrast, there are a range of models in which a ``classical'' gravitational interaction couples to quantized matter, and generally these models predict that masses will not become entangled~\cite{Kafri:2014zsa,tilloy2016sourcing,barvinsky2021correlated,Grossardt:2022zsi,Oppenheim:2018igd,Oppenheim:2023izn,Layton:2023oud,Carney:2023aab}. Thus these experiments can distinguish these two classes of models. We will see below that they can also distinguish the graviton scenario from the entropic scenario, at least in some range of the free parameters of the entropic models. 

Another class of experiments has recently been proposed based on the observation that non-graviton-based models all seem to predict anomalous noise, for example by making the gravitational field into a stochastic variable~\cite{Oppenheim:2018igd,Oppenheim:2023izn,Layton:2023oud}. This is in contrast to standard perturbative quantum gravity mediated by gravitons, which as discussed above produces an intrinsically noiseless (unitary, reversible) interaction. These experiments are thus aimed at directly testing the noise properties of the gravitational interaction~\cite{Kafri:2013wxa,Carney:2021yfw,Oppenheim:2022xjr,Lami:2023gmz}. As we have discussed above, the entropic models have instrinsic noise, and can thus be tested by these experiments.

Here we analyze the predictions of the spin entropic models in both types of experiments. This exercise also allows us to constrain the various free parameters in the models by comparing to existing experiments, as shown in Fig.~\ref{fig:bounds}.

\subsection{Noise in the gravitational force}
\label{sec:noise}

In both the non-local and local entropic models, the gravitational force is generated by a thermalizing mediator and thus has thermally-driven fluctuations. However, the phenomenology of the noise in the two models is different. In the non-local model there is only noise when at least two masses are present, whereas in the local model, even an isolated mass feels noise from the ambient mediator qubits. 

Under the noisy evolution described by Eq.~\eqref{eq:lindblad}, a momentum operator $\mb{p}$ will develop a non-zero variance in its momentum:
\begin{align}
\begin{split}
\label{eq:dp2dt}
\frac{d}{dt} \braket{\mb{p}^2} & = \sum_{\alpha,\pm} \braket{ K_{\alpha,\pm}^\dag \mb{p}^2 K_{\alpha,\pm} } - \frac{1}{2} \braket{ \{ K_{\alpha,\pm}^\dag K_{\alpha,\pm}, \mb{p}^2 \} } \\
& = \sum_{\alpha,\pm} \braket{ \left( \nabla K_{\alpha,\pm} \right)^2}.
\end{split}
\end{align}
To get the second line, we used some commutator algebra and the reality of $K_{\alpha,\pm} = K_{\alpha,\pm}(\mb{x})$. The implication of Eq.~\eqref{eq:dp2dt} is somewhat different in the non-local and local models, as discussed above. In the non-local model, $\mb{p} = \mu (\mb{p}_i/m_i - \mb{p}_j/m_j)$ is the relative momentum between a pair of masses $m_i$ and $m_j$, while in the local model, $\mb{p} = \mb{p}_i$ is the momentum of a given mass $m_i$. Either way, what Eq.~\eqref{eq:dp2dt} shows is that the gravitational force itself is noisy. To compare to experiments, we now turn to a quantitative estimate of this force noise. 

\textbf{Non-local model:} First consider the non-local model, and suppose that we have two masses $m_1$, $m_2$ initially separated by some distance $d$, as in the left panel of Fig.~\ref{fig:experiments}. Inserting the explicit noise operators from Eq.~\eqref{eq:Kpm}, we can easily solve Eq.~\eqref{eq:dp2dt} to find a linear growth in the variance:
\be
\label{eq:dp2-nonlocal}
\braket{\mb{p}^2(t)} = \braket{\mb{p}^2(0)} + S_{FF}(0) t.
\ee
The coefficient $S_{FF}(0) = S_{FF}^{+}(0) + S_{FF}^{-}(0)$ has contributions from both the $K_{\alpha,\pm}$ operators,
\begin{align}
\begin{split}
S_{FF}^{+}(0) & = \frac{G_N m_1 m_2}{d^3(1+\lambda d/\ell^2)} \zeta \mathcal{I}_+ \\
S_{FF}^{-}(0) & =  \frac{G_N m_1 m_2}{d^3(1+\lambda d/\ell^2)} \frac{\mathcal{I}_-}{\zeta},
\end{split}
\end{align}
with 
\begin{align}
\begin{split}
\label{eq:I_pm}
{\cal I}_{+} & = \frac{24}{\pi^2} \int_{0}^{\infty}d\nu \, \nu^{2}g_+(\nu) = \frac{8 +3\pi^2}{32} \approx  1.17 \\
{\cal I}_{-} & = \frac{24}{\pi^2} \int_{0}^{\infty}d\nu \, \nu^{2} g_-(\nu) \approx 1.21.
\end{split}
\end{align}
The linear $t$ scaling in Eq.~\eqref{eq:dp2-nonlocal} is characteristic of Brownian motion, which is why we use the notation of a force noise power at zero frequency $S_{FF}(0)$. This reflects our adiabatic approximation used to derive the Lindblad form of Eq.~\eqref{eq:lindblad}, which applies to observations on timescales long compared to the mediator self-thermalization rate.\footnote{In the long time, low frequency regime that we are considering for heavy masses, we note that standard Brownian motion also predicts a small amount of friction. Concretely, the fluctuation-dissipation theorem predicts $S_{FF}(0) = 4 \mu \gamma_p T$, where $\mu = m_1 m_1/(m_1+m_2)$ is the reduced mass and $\gamma_p$ is the friction term, e.g., $\dot{p} \sim - \gamma_p p$. This gives $\gamma_p \sim G_N (m_1 + m_2)/T d^3(1+\lambda d/\ell^2)$. This can be used to set an independent lower bound on the temperature $T$ of the model at fixed values of $\lambda,\ell$ by requiring, for example, stable planetary orbits over galactic time scales~\cite{feynman2018feynman}. Our adiabatic elimination procedure should recover this friction once kinetic energy terms are included.}

\begin{figure*}[t]
\centering
\includegraphics[width=0.47\linewidth]{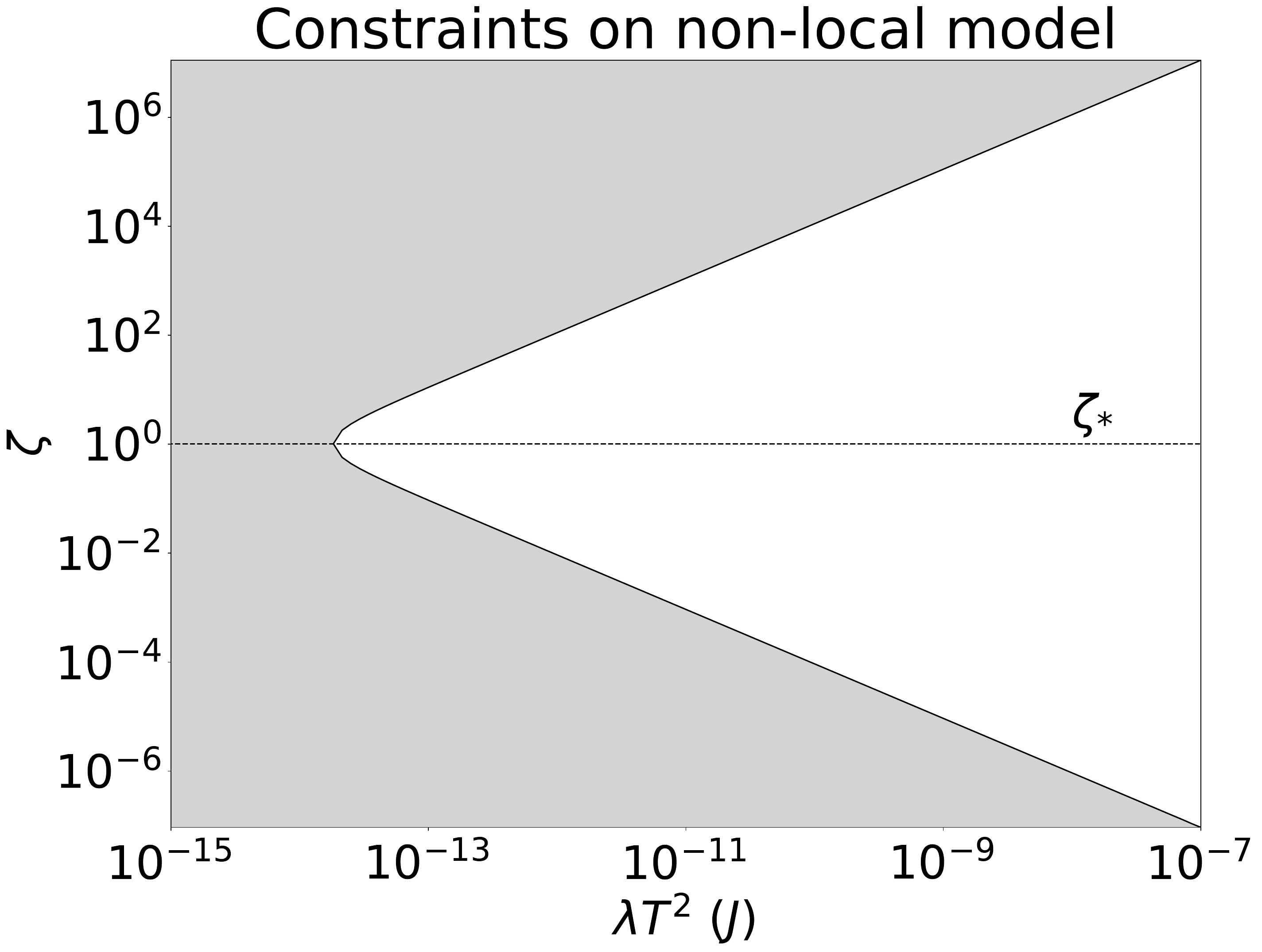}
\includegraphics[width=0.47\linewidth]{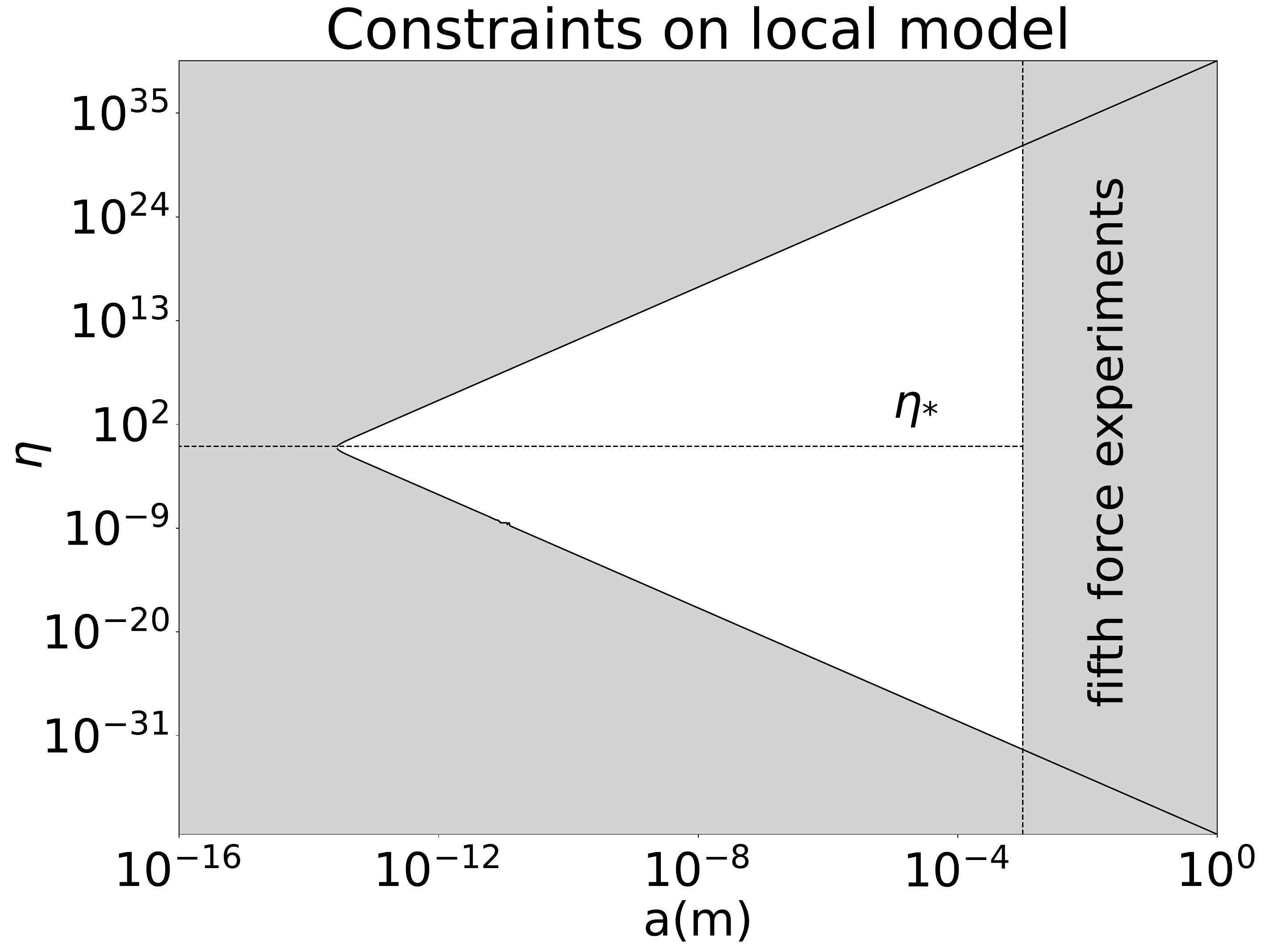}
\caption{\textbf{Experimental constraints on the free parameters.} Left: the local model. The shaded region indicates parameters $\zeta, \lambda T^2$ which are excluded by experiments using cesium atoms ($m_{\rm Cs} \approx 133~{\rm GeV}$) held in spatial superposition at a scale $\delta x \approx 5~{\rm \mu m}$ above the Earth, with observed coherence times of order seconds \cite{kovachy2015quantum,asenbaum2017phase}, i.e., values of $\zeta, \lambda T^2$ which predict a minimum decoherence above $1~{\rm Hz}$ with Eq.~\eqref{eq:deco-nonlocal}. Right: The shaded region indicates parameters $\eta = T(1-\sigma _{*})/\gamma_{th}$, $a$ which are excluded by experiments where objects of $m\sim 1~{\rm kg}$ are held at cryogenic temperatures in a dilution refrigerator with cooling power $\sim 10~{\rm \mu W}$, using Eq.~\eqref{anom_heating}. The shaded region at large lattice spacing $a \gtrsim 1~{\rm mm}$ is excluded by fifth force experiments~\cite{Adelberger:2003zx}, since large values of $a$ indicate deviations from the inverse square law in the local model, as in Eq.~\eqref{eq:f_local_dfn}.}
\label{fig:bounds}
\end{figure*}

This force noise depends on the free parameters $\zeta$ and $\lambda$. The overall noise can be decreased arbitrarily by increasing $\lambda$, or more precisely, by assuming that
\be
\label{eq:lambda-conversion}
\frac{\lambda d}{\ell^2} \gg 1 \iff \frac{\pi^2}{12} \lambda T^2 \gg \frac{G_N m_1 m_2}{d} = U_{\rm grav}.
\ee
Here, we used the emergent $G_N$ in Eq.~\eqref{eq:T-G-many}. Moreover, since the terms $S_{FF}^{\pm}$ enter as a sum of terms $\sim \zeta$ and $\sim 1/\zeta$, the dependence on $\zeta$ takes a minimum value when $\zeta = \zeta_* = \sqrt{{\cal I}_-/{\cal I}_+}$. At this value, we have
\begin{align}
\sqrt{S_{FF}^{\rm min}(0)} \approx 10^{-24}~\frac{{\rm N}}{\sqrt{\rm Hz}} \times \left( \frac{m}{1~{\rm mg}} \right) \left( \frac{1~{\rm mm}}{d} \right)^{3/2}.
\end{align}
To get this numerical value, we took $\lambda = 0$, and benchmarked the experimental parameters based on a class of optomechanical experiments measuring gravity between two mg-scale resonators at mm distance~\cite{westphal2021measurement}. This level of noise is nearly observable with modern ultra-sensitive force measurements. Such a force noise measurement could set a lower bound on $\lambda T^2$, as shown in Fig.~\ref{fig:bounds}.

\textbf{Local model:} In the local model, even an isolated mass feels a noisy background force from thermal fluctuations of the local mediator qubits. Consider a mass $m$ in isolation. According to Eq.~\eqref{eq:dp2dt}, and using the explicit Lindblad operators in Eq.~\eqref{eq:lindblad-op-local}, this mass's momentum variance will again grow linearly in time, according to
\begin{align}
\braket{\mb{p}^2(t)} = \braket{\mb{p}^2(0)} + 2 m t P_{\rm anom}.
\end{align}
The notation $P_{\rm anom} = dE/dt_{\rm anom}$ reflects that this can be viewed as an anomalous heating, with rate
\be
\label{anom_heating}
P_{\rm anom} = P_{+1} + P_{-1},
\ee
where
\begin{align}
\begin{split}
P_{+1} & = \frac{G_N m}{64 \pi a^3} \frac{\gamma_{\rm th}}{T(1-\sigma_*)} \\
P_{-1} & = \frac{G_N m}{8 \pi a^3} \frac{T(1-\sigma_*)}{\gamma_{\rm th}}.
\end{split}
\end{align}
Here we have assumed that the mass is well-localized to some position $\mb{x}$, and approximated the sum over sites $\sum_{\alpha} = \int d^3\mb{r}/a^3$ as a continuum, as above. 

Much like the non-local random force above, the two contributions $P_{\pm}$ add as an inverse combination of the free parameter ratio $\eta=T(1-\sigma_*)/\gamma_{\rm th}$. Thus there is a minimum amount of anomalous heating $P_{\rm min}$,
\begin{align}
\begin{split}
P_{\rm min} & = \frac{G_N m}{8 \sqrt{2} \pi a^3} \\
& \approx  1.3 \times 10^{-20}~{\rm \mu W} \times \left( \frac{m}{1~{\rm GeV}} \right) \left( \frac{10^{-15}~{\rm m}}{a} \right)^3.
\end{split}
\end{align}
Unlike the non-local model, there is no free parameter here that can be tuned to reduce this noise. This provides a very stringent lower bound on $a$. Consider experiments where objects of order $m \sim 1~{\rm kg} \approx 5.6 \times 10^{26}~{\rm GeV}$ are held at cryogenic temperatures in a dilution refrigerator with cooling power on the order of $10~{\rm \mu W}$. This sets a lower bound $a \gtrsim 10^{-13}~{\rm m}$, as displayed in Fig.~\ref{fig:bounds}.

\subsection{Decoherence of spatial superpositions}
\label{sec:decoherence}

One implication of the noisy entropic force is that it causes objects superposed in space to decohere. Consider a density matrix element of the form
\be
\rho(\mb{x},\mb{x}') = \braket{ \mb{x} | \rho | \mb{x}'}.
\ee
In the non-local model, $\mb{x} = \mb{x}_i - \mb{x}_j$ represents the relative position of two masses $m_i$ and $m_j$, whereas in the local model, $\mb{x} = \mb{x}_i$ represents the position of a single mass $m_i$. The noise terms in the Lindblad evolution, Eq.~\eqref{eq:lindblad-final}, give
\be
\dot{\rho}(\mb{x},\mb{x}') = - \frac{1}{2} \sum_{\alpha,\pm} \left( K_{\alpha,\pm}(\mb{x}) - K_{\alpha,\pm}(\mb{x}') \right)^2 \rho(\mb{x},\mb{x}'),
\ee
where we are neglecting the kinetic energy and Newtonian interaction. The solution is
\be
\label{eq:decay-1}
\rho(\mb{x},\mb{x}',t) = e^{-\Gamma(\mb{x},\mb{x}') t} \rho(\mb{x},\mb{x}',0)
\ee
where the decoherence rate $\Gamma = \Gamma_+ + \Gamma_-$, with
\begin{align}
\begin{split}
\label{eq:decay-2}
\Gamma_{\pm} & = \frac{1}{2} \sum_{\alpha} \left( K_{\alpha,\pm}(\mb{x}) - K_{\alpha,\pm}(\mb{x}') \right)^2 \\
& \approx \frac{1}{2} \sum_{\alpha} \left( \nabla K_{\alpha,\pm}(\mb{x}) \cdot \delta \mb{x} \right)^2.
\end{split}
\end{align}
In the last line, the approximation is for a small superposition, where we can expand $\mb{x}' = \mb{x} + \delta \mb{x}$. This is clearly related to the force noise in Eq.~\eqref{eq:dp2dt}. Let us now turn to some quantitative estimates of this effect in the two models.

\textbf{Non-local model:} Consider two masses $m_1$ and $m_2$. Suppose that initially one mass is held in a fixed location, and the other is superposed, 
\be
\ket{\psi} = \frac{1}{\sqrt{2}} \ket{ \mb{x}_1} \otimes \left( \ket{\mb{x}_2} + \ket{\mb{x}'_2} \right).
\ee
See Fig.~\ref{fig:experiments}, center. The relative coordinate $\mb{x}$ is thus similarly superposed. Let $\mb{x}_1 - \mb{x}_2 = \mb{d}$ and $\mb{x}_1 - \mb{x}_2' = \mb{d} + \delta \mb{x}$, where in a typical realistic experiment, $\delta x \ll d$. According to Eq.~\eqref{eq:decay-1}, the superposition will decohere. Continuing to neglect the effects of kinetic energy and the coherent Newtonian force, the center-of-mass state evolves trivially, and Eq.~\eqref{eq:decay-2} predicts that the $m_2$ state will decohere with rate $\Gamma = \Gamma_+ + \Gamma_-$, where
\begin{align}
\begin{split}
\label{eq:deco-nonlocal}
\Gamma_{+} & \approx \frac{G_N m_{1}m_{2} \delta x^2 \cos^2\theta}{2d^3(1 + \lambda d/\ell^2)} \zeta \mathcal{I}_{+} \\
\Gamma_{-} & \approx \frac{G_{N}m_{1}m_{2} \delta x^2 \cos^{2}\theta}{2d^{3}(1+ \lambda d/\ell^2)} \frac{{\cal I}_{-}}{\zeta}.
\end{split}
\end{align}
Here the dimensionless integrals ${\cal I}_{\pm}$ were given above in Eq.~\eqref{eq:I_pm}, and $\theta$ is the angle between $\mb{d}$ and $\delta \mb{x}$. Again, the free parameter $\lambda d/\ell^2$ in the denominator is related to the masses through Eq.~\eqref{eq:lambda-conversion}.

Much like the force noise in the previous section, we see that the decoherence rate has contributions depending both on the damping parameter $\zeta$ and $1/\zeta$. At a given value of the free parameter $\lambda$, there is thus a minimum decoherence rate: any superposed mass will decohere on a time scale at least as fast as 
\begin{align}
\begin{split}
\label{eq:gamma-min}
& \Gamma_{\rm min} \approx  \frac{G_N m_{1}m_{2} \delta x^2}{d^3(1 + \lambda d/\ell^2)}\sqrt{{\cal I}_+ {\cal I}_-} \\
& \approx 1~{\rm kHz} \times \left( \frac{m_1}{M_{\rm Earth}} \right) \left( \frac{m_2}{m_{\rm Cs}} \right) \left( \frac{R_{\rm Earth}}{d} \right)^3 \left( \frac{\delta x}{0.5~{\rm m}} \right)^2.
\end{split}
\end{align}
In the second line, we took $\lambda = 0$. However, since the combination $\lambda /\ell^{2}$ appears in the denominator, we can reduce the decoherence rate without affecting the emergent of Newtonian force by choosing $\lambda/\ell^{2} \gg 1/d$. 

Equation \eqref{eq:gamma-min} can be compared with experiment, and sets a constraint on the free parameter $\lambda/\ell^2$. The numerical values in the second line are based on experiments with atomic interferometers on the surface of Earth, where a cloud of atoms is superposed by as much as $\delta x \approx 0.5~{\rm m}$~\cite{kovachy2015quantum,asenbaum2017phase}. Future experiments are targeting $\delta x \gtrsim 100~{\rm m}$~\cite{badurina2020aion,abe2021matter}. The results of the existing experiments are consistent with cesium atoms ($m_{\rm Cs} \sim 133~{\rm GeV}$) remaining in coherent superposition over times of order seconds, so that $\Gamma_{\rm min} \gtrsim 1~{\rm Hz}$ is excluded. This is already enough to completely rule out the model with $\lambda = 0$. We show the general numerical constraints in Fig.~\ref{fig:bounds}.

\textbf{Local model:} Now consider the local model. Suppose we have a single massive object placed in spatial superposition of two locations $\ket{\psi} = (\ket{\mb{x}} + \ket{\mb{x}'})/\sqrt{2}$. Just like in the non-local model, this superposition will decohere due to the interactions of the mass with the mediator qubits. Unlike the non-local model, this happens regardless of whether there are other masses present.

The decoherence rate can be estimated similarly to the non-local model. From Eqs.~\eqref{eq:decay-1} and \eqref{eq:decay-2}, we see that the relevant off-diagonal density matrix element will decay with rate $\Gamma = \Gamma_+ + \Gamma_-$, with
\begin{align}
\begin{split}
\label{eq:deco_rates_local}
\Gamma_{+} & = G_N m^2 \frac{1}{4\pi^2} \frac{\gamma_{\rm th}}{T(1-\sigma_*)} \left[ \frac{1}{2a} - \frac{2\pi}{3 \delta x} + \cdots \right] \\
\Gamma_{-} & = G_N m^2 \frac{2}{\pi^2} \frac{T(1-\sigma_*)}{\gamma_{\rm th}} \left[ \frac{1}{2a} - \frac{2\pi}{3 \delta x} + \cdots \right].
\end{split}
\end{align}
Here, we used the explicit Lindblad operators in Eq.~\eqref{eq:lindblad-op-local}, took the usual continuum limit, and assumed that $\mb{x}' = \mb{x} + \delta \mb{x}$ with $a/\delta x \ll 1$.

The two terms $\Gamma_{\pm}$ depend on the free parameters of the model only through the ratio $\eta = T(1-\sigma_*)/\gamma_{\rm th}$, with the same kind of inverse relationship we have seen before. This implies that there is a minimal decoherence rate predicted by the model, when $\eta = \eta_* = 2 \sqrt{2}$:
\be
\label{eq:min-decay-local}
\Gamma_{\rm min} = \frac{G_N m^2}{\sqrt{2} \pi^2 a} \approx 2.5~{\rm Hz} \times \left( \frac{m}{m_{\rm Cs}} \right)^2 \left( \frac{10^{-27}~{\rm m}}{a} \right).
\ee
This sets a lower bound on possible values of $a$. Here we used the same cesium atom interferometer numbers from the previous section, which give a bound $a \gtrsim 10^{-27}~{\rm m}$. This is a much weaker bound than that obtained from anomalous heating in Sec.~\ref{sec:noise}, because here the noise $\sim 1/a$, whereas the force noise $\sim 1/a^3$.

\subsection{Entanglement generation between masses}
\label{sec:entanglement}

Finally, consider the experiment shown in the right panel of Fig.~\ref{fig:experiments}, which aims to test whether gravitational interactions can entangle two massive bodies~\cite{bose2017spin}. Many alternative versions of the same basic test have also been proposed~\cite{marletto2017gravitationally,Matsumura:2020law,Datta:2021ywm,Lami:2023gmz}, and we expect that our conclusions here generalize to those settings in a straightforward way. 

Concretely, suppose we have two objects with the same mass $m$ freely falling and separated appropriately so that their interactions are dominated by their mutual gravity. We further choose a geometry so that the $\ket{R}_1 \ket{L}_2$ path pair is short compared to the other three pairs. We will also continue to assume that we can neglect the kinetic energy $\mb{p}^2/2m$ terms relative to the gravitational interaction.

We prepare the two masses in an initial product state of the form
\be
\label{eq:psi0}
\ket{\psi(0)} = \frac{1}{2} \left( \ket{L} + \ket{R} \right) \otimes \left( \ket{L} + \ket{R} \right),
\ee
which is unentangled. We assume the position widths of the various $\ket{L,R}$ states are narrow enough that they have no overlap,  $\braket{L|R} = 0$. The question is whether time evolution under the gravitational interaction causes this state to become entangled. To quantify this, consider the observable~\cite{bose2017spin}
\be
W = X_1 \otimes Z_2 + Y_1 \otimes Y_2
\ee
where these are Pauli operators in the pseudospin basis $\ket{0} := \ket{L}$, $\ket{1} := \ket{R}$. This operator is known as an entanglement witness. It can be easily shown that $|\braket{W}| > 1$ can only occur in entangled states~\cite{terhal2000bell}. Thus if we initialize the system in a product state, time evolve, and measure $|\braket{W}| > 1$, then we know the gravitational interaction generated entanglement. 

\begin{figure}[t]
\includegraphics[width=0.95\linewidth]{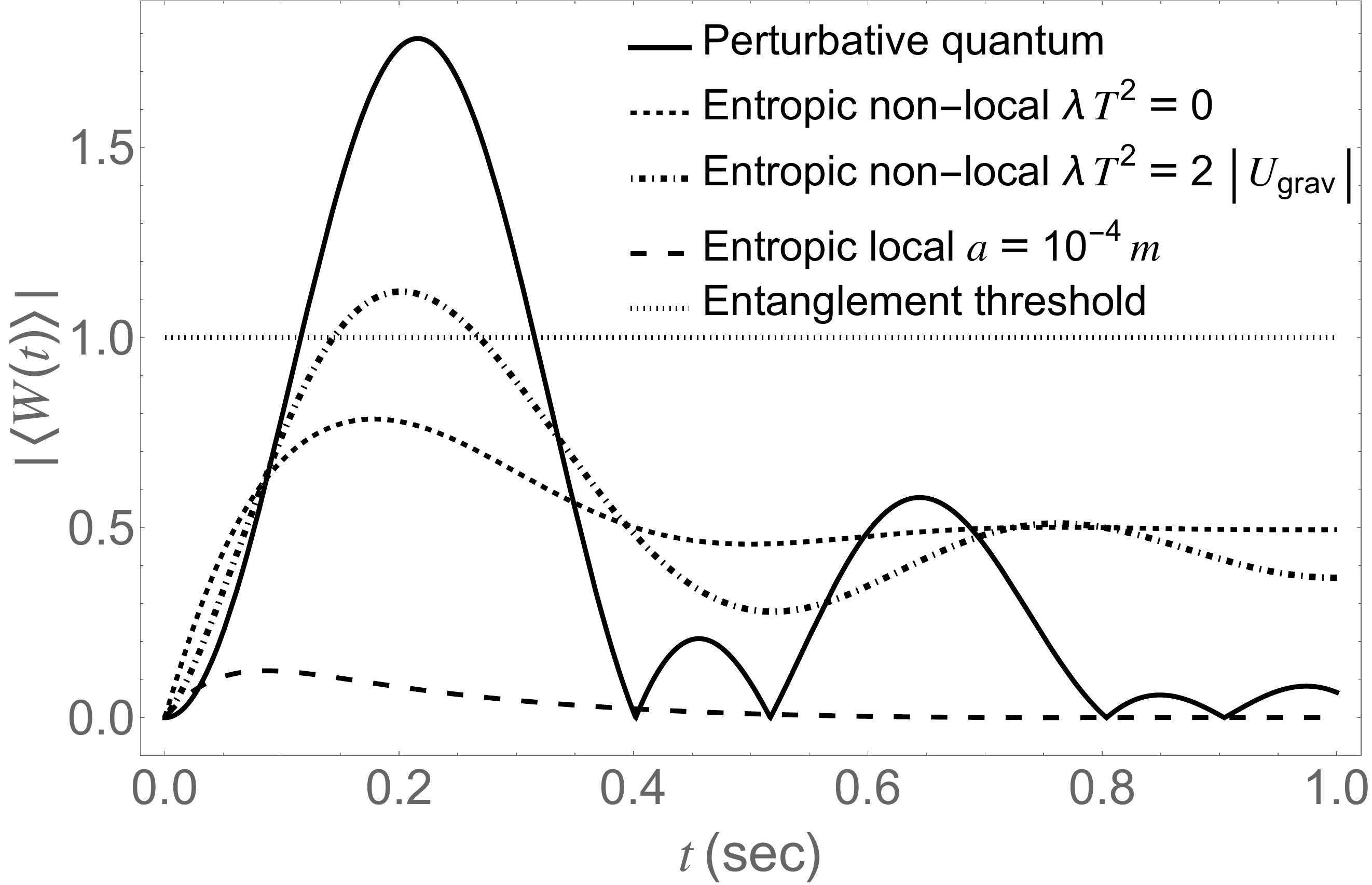}
\caption{Behavior of the entanglement witness $\left| \braket{W(t)}\right|$ as a function of time in both standard perturbative quantum gravity and our entropic gravity models. Values above the threshold $\left| \braket{W}\right| > 1$ represent states which are provably entangled. We used the following values for the plots shown above: $m_{1}=m_{2}=10^{-13}\,\rm kg$, $\delta x=250\, \mu \rm m$, $d=450\, \mu \rm m$~\cite{bose2017spin} and $a=10^{-4}\,m$. For the perturbative quantum plot we used Eq.~\eqref{W_per_quantum}, for the non-local Eq.~\eqref{W_entropic_non_local}, and for the local Eq.~\eqref{W_entropic_local}, both with minimal decoherence. We see that for non-zero $\lambda T^2= 2\left|U_{\rm grav}\right|$ with $U_{\rm grav}=G_{N}m^2/d$, the decoherence in the non-local is reduced and a tiny short-lived entangled state is produced. For the local-model we find that no entanglement is produced even for very large value of the lattice spacing $a=10^{-4}\,m$.}
\label{fig:W}
\end{figure}

For a general density matrix in the $\ket{L}$,$\ket{R}$ basis for the two masses, a short calculation shows that the expectation value of $W$ is
\begin{align}
\label{eq:Wt}
\braket{W} = 2 {\rm Re} \Big[ \rho_{LL,RL} - \rho_{LL,RR}  + \rho_{LR,RL} - \rho_{LR,RR} \Big].
\end{align}
The question is then how this entanglement witness evolves in various models of gravity. In Fig.~\ref{fig:W}, we show the behavior of the entanglement witness $\left|\braket{W}\right|$ as a function of time in both perturbative quantum gravity and our entropic models.

In ordinary perturbative quantum gravity, we have unitary time evolution generated by just the Newtonian potential operator, with no noise. The initial product state $\ket{\psi(0)}$ evolves into an entangled state of the form
\be
\label{eq:psi-t}
\ket{\psi(\Delta t)} = \frac{1}{2} \left( \ket{LL} + e^{i \phi_{LR}} \ket{LR} + e^{i \phi_{RL}} \ket{RL} + \ket{RR} \right)
\ee
up to an overall phase. The differential phases $\phi_{LR}$, $\phi_{RL}$ are
\be
\label{eq:phi-grav}
\begin{split}
\phi_{LR} &\approx -\frac{G_N m^2 \delta x \Delta t}{d^2}\left(1- \frac{\delta x}{d}\right), \\ \phi_{RL} &\approx\frac{G_N m^2 \delta x \Delta t}{d^2}\left(1+ \frac{\delta x}{d}\right),
\end{split}
\ee
where $\Delta t$ is the free-fall time\footnote{More realistically, $\Delta t$ is the time until the first decohering error occurs, e.g., the time at which an ambient gas molecule hits one of the masses or one of the masses emits a blackbody photon. We will ignore these effects here for simplicity, since our goal is to distinguish the purely gravitational effects.} and we have only kept terms up to ${\cal O}((\delta x/d)^{2})$. The state $\ket{\psi(\Delta t)}$ is entangled after sufficiently long times. To be more precise, it has density matrix elements
\begin{align}
\begin{split}
\rho_{LL,RL} & = \frac{1}{4} e^{-i \phi_{RL}}, \ \ \ \rho_{LR,RR} = \frac{1}{4} e^{i \phi_{LR}} \\
\rho_{LR,RL} & = \frac{1}{4} e^{i (\phi_{RL}-\phi_{LR})}, \ \ \ \rho_{LL,RR} = \frac{1}{4}.
\end{split}
\end{align}
Using this in Eq.~\eqref{eq:Wt}, we have
\begin{align}
\label{W_per_quantum}
\left| \braket{W} \right|= \frac{1}{2}\left| \cos(\phi_{LR}-\phi_{RL})+\cos\phi_{RL} - \cos \phi_{LR}-1\right|,
\end{align}
so in particular $\left| \braket{W} \right| > 1$ 
after the gravitational interaction has acted over a sufficiently long time. Thus perturbative quantum gravity is observably and provably entangling.

The situation in our entropic models is more subtle. Since we have an open system governed by the Lindbladian evolution of Eq.~\eqref{eq:lindblad-final}, the initial state $\ket{\psi(0)}$ now evolves into a mixed state $\rho(\Delta t)$. This mixed state can still, in principle, have entanglement. In particular, the effective Newtonian interaction in Eq.~\eqref{eq:lindblad-final} can generate entanglement in the usual sense. The question is to what extent the noise terms destroy this entanglement.

\textbf{Non-local model:} In the non-local model, the same density matrix elements which encode the entanglement will also decohere. The rate of entanglement is the same as the perturbative gravity calculation since it is controlled by the same potential operator, while the decoherence rates were calculated in Sec.~\ref{sec:decoherence}. Putting these together, we find
\begin{align}
\begin{split}
\label{eq:rho-t-us}
\rho_{LL,RL} & = \frac{1}{4} e^{-i \phi_{RL}} e^{-\Gamma \Delta t}, \ \ \ \rho_{LR,RR} = \frac{1}{4} e^{-i \phi_{LR}} e^{-\Gamma \Delta t} \\
\rho_{LR,RL} & = \frac{1}{4} e^{i (\phi_{RL}-\phi_{LR})} e^{-4\Gamma \Delta t}, \ \ \ \rho_{LL,RR} = \frac{1}{4}.
\end{split}
\end{align}
The explicit decoherence rate $\Gamma$ in the non-local model is given above in Eq.~\eqref{eq:deco-nonlocal}. The $4$ in the exponent on the second line is geometrical, see Fig.~\ref{fig:experiments}. In total, this gives the entanglement witness evolution
\bes{
\label{W_entropic_non_local}
\left| \braket{W} \right| = \frac{1}{2}\big|&e^{-4\Gamma \Delta t}\cos(\phi_{LR}-\phi_{RL}) \\&+ e^{-\Gamma \Delta t}(\cos\phi_{RL} -\cos \phi_{LR}) -1\big|.
}
We see the competition between the entangling and decohering effects. This competition is tuneable in the sense that the entanglement is independent of the free parameter $\lambda$, while the decoherence $\Gamma \to 0$ as $\lambda T^2 \to \infty$. In Fig.~\ref{fig:W}, we see that at $\lambda=0$, there is no observable amount of entanglement, whereas at sufficiently large $\lambda$, the decoherence can be totally removed and the model becomes indistinguishable from standard perturbative gravity. If we use the minimum value $(\lambda T^{2})_{\rm min}\sim 10^{-14} \rm J$ consistent with current atom interferometers, as shown in Fig.~\ref{fig:bounds}, the decoherence is negligible. 

\textbf{Local model:}
In the local model, all the terms appearing in Eq. \eqref{eq:Wt} evolve. The reason is that now the decoherence is local, and so even a density matrix element like $\ket{LL}\bra{RR}$ decoheres, even though the there is no gravitational phase generated. In general, the density matrix elements evolve as
\begin{align}
\begin{split}
\label{eq:local_dm_elements_main}
\rho_{LL,RL} & = \frac{1}{4} e^{-i \phi_{RL}} e^{-\Gamma_{0} \Delta t},~\rho_{LR,RR}=\frac{1}{4} e^{i \phi_{LR}} e^{- \Gamma_{0} \Delta t},\\
\rho_{LR,RL} & =\frac{1}{4} e^{i(\phi_{RL}-\phi_{LR})} e^{-\Gamma_{+} \Delta t},~\rho_{LL,RR}=\frac{1}{4}e^{-\Gamma_{-} t}.
\end{split}
\end{align}
The witness evolves as 
\begin{align}
\begin{split}
\label{W_entropic_local}
\left| \braket{W} \right| & = \frac{1}{2}\Big| e^{-\Gamma_{+} \Delta t}\cos(\phi_{LR}-\phi_{RL})  \\ 
& +e^{- \Gamma_{0} \Delta t}\left(\cos\phi_{RL} - \cos \phi_{LR}\right) -e^{-\Gamma_{-} \Delta t} \Big|. 
\end{split}
\end{align}
In the limit that the superpositions are small compared to the lattice and average spacing, these decoherence rates are given by Eq.~\eqref{eq:deco_rates_local}. When the superposition scale or lattice size is comparable to the distance between the masses $d$, the expressions are more complicated. Since this latter regime was nominally proposed in \cite{bose2017spin}, we record the full expressions in Appendix~\ref{app:witness}.

In Fig.~\ref{fig:W}, we plot $\left|\braket{W}\right|$ for the local model. Even with the minimal decoherence rate consistent with the anomalous heating bounds of Sec.~\ref{sec:noise}, we find that decoherence in the local model significantly reduces the witness below the entanglement threshold. Thus, any detection of gravitational entanglement in such an experiment would rule out the entire remaining parameter space of the local spin entropic model.

\section{Outlook}

The gravitational interactions we observe at accessible length scales could in principle emerge in many ways from physics at the Planck scale $\rho \sim m_{\rm Pl}/\ell_{\rm Pl}^3 \sim 10^{104}~{\rm J}/{\rm cm}^3$. Perhaps the simplest is that gravitational perturbations are quantized as gravitons, i.e., as another quantum field theory like the gauge bosons of the other fundamental forces in nature. This is a perfectly good effective quantum field theory; nothing in principle forces us to abandon this picture until energies near the Planck scale~\cite{Donoghue:1994dn,Burgess:2003jk,Donoghue:2022eay}. However, it is interesting to consider alternative possibilities, particularly given that we may be able to constrain these with experiments. 

Here we have presented a phenomenology in which gravity emerges as an entropic effect mediated by a many-body system, building on the pioneering ideas of~\cite{jacobson1995thermodynamics,verlinde2011origin}. This picture is radically different from the simple graviton scenario both in ontology and, at least for some values of the free parameters defining the models, in its predictions. However, we have also found that for certain values of the free parameters (e.g., for $\lambda \to \infty$ in the non-local model of Sec.~\ref{sec:general-construction}), this kind of entropic interaction can also reproduce the predictions of ordinary virtual graviton exchange, at least in the non relativistic limit studied here. It would be very interesting to understand if there is some fundamental upper bound on $\lambda$, but we have not been able to identify one.

While we have presented a specific set of models based on thermalizing spins for concreteness, we expect that the qualitative phenomenology found in this paper generalizes to a wider class of ``entropic'' models of gravity. In particular, the essential idea here is that if an interaction is produced by some thermal mediator system, it should generate thermal fluctuations in the force, which are observable. This conclusion seems to hold quite generally. In particular, we still find such fluctuations even at the special point the local spin model of Sec.~\ref{sec:local-construction} in which the force is purely entropic and has no internal energy fluctuations. Of course, fully mapping out the generality of these conclusions is a crucial open question, particularly in the relativistic context.

Within the broader scope of models which have been constructed to produce observables that differ from standard graviton physics~\cite{tilloy2016sourcing,barvinsky2021correlated,Grossardt:2022zsi,Oppenheim:2018igd,Oppenheim:2023izn,Layton:2023oud,Carney:2023aab,Oppenheim:2018igd,Oppenheim:2023izn,Layton:2023oud}, the model here has the substantial advantage that it is subject to the usual rules of quantum mechanics: unitarity is fundamentally valid at the microscopic scale, the evolution in the state is linear, and there are no fundamental classical stochastic variables. In particular, one should be able to study renormalization in this model using standard tools, which we leave for future work.

The directions moving forward are clear. One is to eliminate the crutch of a thermal reservoir, by replacing the thermalizing oscillators and their bath by some simpler self-thermalizing system. Another is to find a relativistic version of the construction presented here. And the third, by far the most important, is to continue the march toward experimental tests of these scenarios.

\section*{Acknowledgments}

We thank Joshua Batson, Yanbei Chen, Patrick Draper, Chris Jarzynski, Mark Mitchison, Jess Riedel, Mark van Raamsdonk, Erik Verlinde, and Nicole Yunger Halpern for helpful discussions, Tehya Andersen and Jordan Wilson-Gerow for collaboration on early versions of these ideas, and the NSF-funded Harvard ITAMP and Aspen Center for Physics for enabling fruitful conversations during visits. Work at Berkeley Lab is supported by the U.S. Department of Energy, Office of High Energy Physics, under Contract No. DEAC02-05CH11231 and by DOE Quantum Information Science Enabled Discovery (QuantISED) for High Energy Physics grant KA2401032. This work was made possible
by the Heising-Simons Foundation grant 2023-4467 ``Testing the Quantum Coherence of Gravity''. J. T. is solely funded by the National Institute of Standards and Technology.

\bibliographystyle{utphys-dan}
\bibliography{entropic-forces}
\clearpage

\newpage

\appendix

\section{Oscillator mediators}
\label{appendix:oscillators}

In the main text, we presented models with a mediator system composed of qubits. As mentioned around Eqs.~\eqref{eq:H-first} and \eqref{eq:omega-f}, an analogous construction where the qubits are replaced by oscillators can also produce an emergent gravitational force, at least in the non-local version of the model. Here we report the basic setup to highlight the generality of the ideas presented in the main text.

Instead of the qubit-mass Hamiltonian of Eq.~\eqref{eq:H-first}, consider a pair of masses coupled to a system of harmonic oscillators labeled by $\alpha=1,2,\ldots$. We use the same kind of interaction, where now the oscillator frequencies are dependent on the relative position of the two masses:
\be
\label{eq:H-osc}
H = \sum_{i=1,2} \frac{\mb{p}_i^2}{2 m_i} - \sum_{\alpha=1}^{\infty} \omega_{\alpha}(\mb{x}) N_{\alpha}, \ \ \ N_{\alpha} = a_{\alpha}^\dag a_{\alpha}.
\ee
Here $a_{\alpha}$ is the annihilation operator of the $\alpha$th oscillator. The mediator frequencies are taken to have a massless (linear) spectrum identical to that in Eq.~\eqref{eq:omega-f}, viz., $\omega_{\alpha}(\mb{x}) = \alpha f(\mb{x})$. 

We follow the discussion in the main text. First consider the limit where the masses can be treated classically, so $\mb{x}$ is a c-number. Let the oscillators be prepared in the thermal state,
\be
\label{eq:app-rhoT}
\rho_T = \bigotimes_{\alpha=1}^{\infty} \rho_{T,\alpha,\mb{x}}, \ \ \rho_{T,\alpha,\mb{x}} = \sum_{m=0}^{\infty} \frac{e^{-m \omega_{\alpha}(\mb{x})/T}}{Z_{T,\alpha,\mb{x}}} \ket{m} \bra{m} ,
\ee
where the single-mode partition function is $Z_{T,\alpha,\mb{x}} = [1-e^{-\omega_{\alpha}(\mb{x})/T}]^{-1}$. The average internal energy of the $\alpha$th oscillator is
\be
U_{\alpha} = \sum_m m \omega_{\alpha}(\mb{x}) \frac{e^{-m \omega_{\alpha}(\mb{x})/T}}{1 - e^{-\omega_{\alpha}(\mb{x})/T}} = \omega_{\alpha}(\mb{x}) \frac{e^{-\omega_{\alpha}(\mb{x})/T}}{1 - e^{-\omega_{\alpha}(\mb{x})/T}}.
\ee
Again we take the continuum limit $\sum_{\alpha} \to \int d\alpha$, and get the total internal energy
\be
U = \int_0^{\infty} \frac{d\omega}{f(\mb{x})}  \frac{\omega e^{-\omega/T}}{1-e^{-\omega/T}} = \frac{\pi^2}{6} \frac{T^2}{f(\mb{x})}.
\ee
Similar manipulations give the entropy:
\be
S = \frac{\pi^2}{3} \frac{T^2}{f(\mb{x})}.
\ee
Assuming that the oscillators remain thermalized at temperature $T$, they will drive the massive bodies with a thermal/entropic force in order to extremize their free energy $\mathcal{A}$. This force is
\be
\label{eq:FN-app}
F_{\rm th} = - \nabla \mathcal{A} = - \frac{\pi^2}{6} T^2 \frac{\nabla f(\mb{x})}{f^2(\mb{x})}.
\ee
This is the same formula as we obtained with the qubit mediators, up to a $1/2$ which comes from the use of the Boltzmann rather than Fermi-Dirac distribution. Thus we can again choose the oscillator frequencies to have dependence on the mass positions in the form $f(\mb{x}) = |\mb{x}|/\ell^2$, where $\ell$ is some constant with dimensions of length, and obtain
\be
\label{eq:app-Fth}
F_{\rm th} = -G_N m_1 m_2 \frac{\hat{\mb{x}}}{|\mb{x}|^2}.
\ee
Once again, to match to Newton's law, we make the identification
\be
\label{eq:app-T-G}
\frac{\pi^2}{6} T^2 \ell^2 \equiv G_N m_1 m_2.
\ee
With this identification, we obtain Newton's law of gravity from the interaction with a collection of thermally-driven oscillators.

Much of the rest of the main text can be repeated in this model, including the microscopic Hamiltonian, calculation of force noise, and so forth. The steps follow the exact same line of reasoning. The only major technical difference is the adiabatic elimination technique discussed in Appendix \ref{app:adiabatic}. In the oscillator case, this becomes an infinite-dimensional linear algebra problem. It can be solved in principle using Laguerre polynomials, but the solution is sufficiently unwieldy that we will omit it here.

\section{Mass and mediator Lindblad equation}
\label{appendix:lindblad}

In this Appendix we review the derivation of the Lindblad equation describing the masses and mediator, Eq.~\eqref{eq:lindblad}. The material is standard, and we follow~\cite{wang2016quantum,Nathan:2020qck}, but include it for completeness and to spell out the details of our bath modeling. 

Consider the full Hamiltonian \eqref{eq:H-full}. In the interaction picture we define 
\be
\begin{split}
V_{SB,I}&= U^{\dagger}_{I}(t)V_{SB}U_{I}(t)\,, \\
\rho_{SB,I}(t)&=U^{\dagger}_{I}(t)\rho_{SB}U_{I}(t)\,,
\end{split}
\ee
where $U_{I}(t)=e^{-i(H_{S}+ H_{B})t}$. The Liouville–von Neumann equation is 
\be
\label{Interaction_LN_equation}
\frac{d}{dt}\hat \rho_{I,SB}(t)=-i\left[\hat V_{SB,I}(t),\hat\rho_{I,SB}(t)\right]\,.
\ee
Integrating from $t_{0}$ to $t$ we obtain (we drop the subscript I from this point on)
\be
\label{density_matrix_SM_eq_t1_t2_2}
\hat \rho_{SB}(t)=\hat\rho_{SB}(t_{0}) -i\int_{t_{0}}^{t}dt'\, \left[\hat V_{SB}(t'),\hat\rho_{SB}(t')\right]\,.
\ee
Substituting \eqref{density_matrix_SM_eq_t1_t2_2} into \eqref{Interaction_LN_equation} and tracing out the bath $\rho_{S}=\tr_{B}\rho_{SB}$ we obtain
\be
\label{System_density_matrix_time_evolution_1_many}
\begin{split}
\frac{d}{dt}\rho_{S}=& -i\tr_{B}\bigg\{\left[ V_{SB}(t),\rho_{SB}(t_{0})\right] \bigg\}\\&- i\int_{t_{0}}^{t}dt' \tr_{B}\bigg\{\left[V_{SB}(t),\left[ V_{SB}(t'), \rho_{SB}(t')\right]\right] \bigg\}\,.
\end{split}
\ee
In the Born approximation $\rho_{SB}(t)\approx \rho_{S}(t)\otimes \rho_{B,T}$ 
where $\rho_{B,T}=Z^{-1}e^{-\frac{H_S}{T}}$.
$\left[ V_{SB}(t),\rho_{SB}(t_{0})\right]=0$ and Eq. \eqref{System_density_matrix_time_evolution_1_many} becomes
\begin{align}
\begin{split}
\frac{d}{dt}\rho_{S}(t) & =  -\int_{t_0}^{t} dt'\bigg[
J_{-}(t-t')\left[\sigma_{+,I}(t),\sigma_{-,I}(t')\rho_{S}(t')\right]\\
& +J_{+}(t-t')\left[\sigma_{-,I}(t),\sigma_{+,I}(t')\rho_{S}(t')\right]+ \text{h.c.}
\bigg]\,,
\label{System_density_matrix_time_evolution_2_many}
\end{split}
\end{align}
where $\sigma_{\pm,I}(t):=U^{\dagger}_{I}(t)\sigma_{\pm}U_{I}(t)$ and $J_{-}(t-t')=\tr_{B}\hat F(t)\hat F^{\dagger}(t')\rho_{B,T}$, $J_{+}(t-t')=\tr_{B}\hat F^{\dagger}(t)\hat F(t')\rho_{B,T}$ and $$\hat F(t)=\sum_{k}\left(ig_{k}\,e^{-if_{k}t}b_{k}\right)\,.$$
For simplicity we assumed $g_{k}=-ig$ with $g\in \mathbb{R}$ and consider a single qubit bath, suppressing the index $\alpha$ in Eq. \eqref{eq:bath-terms} .

In the Markov approximation we make the replacement $\rho_{S}(t')\approx \rho_{S}(t)$ and take the limit $t_{0}\to-\infty$ in Eq. \eqref{System_density_matrix_time_evolution_2_many}. If we  define spectral functions $J_{s}(\omega)=(2\pi)^{-1}\int_{-\infty}^{\infty}dtJ_{s}(t)e^{-is\omega t}$ and jump correlators $G_{s}(\omega)=\sqrt{J_{s}(\omega)/2\pi}$ for $s=\{+,-\}$, and use the convolution property $J_{s}(t-t')=\int_{-\infty}^{\infty}ds G_{i}(t-s)G_{i}(s-t')$ into Eq. \eqref{System_density_matrix_time_evolution_2_many} we obtain
\be
\label{System_density_matrix_time_evolution_2_Markov_convolution_many}
\begin{split}
\frac{d}{dt}\rho_{S}&=\int_{-\infty}^{\infty}dt' \int_{-\infty}^{\infty}ds\, {\cal F}_{-}(t,s,t')[\rho_{S}(t)]\\&
+ \int_{-\infty}^{\infty}dt' \int_{-\infty}^{\infty}ds\, {\cal F}_{+}(t,s,t')[\rho_{S}(t)]\,,
\end{split}
\ee
where
\be
\label{F_plus_minus_def_many}
\begin{split}
&{\cal F}_{-}(t,s,t')[\rho_{S}(t)]=\\&-\theta(t-t')G_{-}(t-s)G_{-}(s-t')\sum_{n} \left[\sigma_{+,I}(t),\sigma_{-,I}(t')\rho_{S}(t)\right]\\&+ \text{h.c.}\,,\\&
{\cal F}_{+}(t,s,t')[\rho_{S}(t)]=\\&- \theta(t-t')G_{+}(t-s)G_{+}(s-t')\sum_{n} \left[\sigma_{-,I}(t),\sigma_{+,I}(t')\rho_{S}(t)\right]\\&+ \text{h.c.}\,.
\end{split}
\ee
Next, we integrate \eqref{System_density_matrix_time_evolution_2_Markov_convolution_many} from $t_{1}$ to $t_{2}$ that will be much longer than the decay scale of the bath correlation functions
\be
\label{System_density_matrix_time_evolution_2_Markov_convolution_integrated_many}
\begin{split}
&\rho_{S}(t_{2})-\rho_{S}(t_{1})\approx\\& \int_{t_{1}}^{t_{2}}dt\,\int_{-\infty}^{\infty}dt' \int_{-\infty}^{\infty}ds\, {\cal F}_{-}(t,s,t')[\rho_{S}(t)]\\&
+ \int_{t_{1}}^{t_{2}}dt\,\int_{-\infty}^{\infty}dt' \int_{-\infty}^{\infty}ds\, {\cal F}_{+}(t,s,t')[\rho_{S}(t)]
\end{split}
\ee
We now make the following approximations~\cite{Nathan:2020qck}. Due to the fast decay of the bath correlation functions we can replace $\rho_{S}(t)$ with $\rho_{S}(s)$ and change the integration domain from $-\infty<(s,t')<\infty$, $t_{1}\le t\le t_{2}$ to the domain $-\infty< (t,t')<\infty$, $t_{1}\le s\le t_{2}$. We obtain
\be
\begin{split}
&\rho_{S}(t_{2})-\rho_{S}(t_{1})\approx\\& \int_{t_{1}}^{t_{2}}ds \int_{-\infty}^{\infty}dt\int_{-\infty}^{\infty}dt' {\cal F}_{-}(t,s,t')[\rho_{S}(s)]\\&+ \int_{t_{1}}^{t_{2}}ds \int_{-\infty}^{\infty}dt\int_{-\infty}^{\infty}dt' {\cal F}_{+}(t,s,t')[\rho_{S}(s)]\,.
\end{split}
\ee
By taking time derivative with $t_{2}$ and renaming the integration variables we get
\be
\label{System_density_matrix_time_evolution_f1_many}
\begin{split}
&\frac{d}{dt}\rho_{S}(t)\approx \int_{-\infty}^{\infty}ds\int_{-\infty}^{\infty}ds' {\cal F}_{-}(s,t,s')[\rho_{S}(t)]\\&
+ \int_{-\infty}^{\infty}ds\int_{-\infty}^{\infty}ds' {\cal F}_{+}(s,t,s')[\rho_{S}(t)]\,.
\end{split}
\ee
Finally, we can assume that operators $\sigma_{\pm,I}(s)$ change slowly within the integrals in  \eqref{System_density_matrix_time_evolution_f1_many} and replace them with $\sigma_{+,I}(s)\to e^{-i \omega(\hat{\mb{x}}_{I}(t))s}\sigma_{-}$   \cite{wang2016quantum}. Going back to the Schr\"{o}dinger picture $\dot \rho_{S} (t)= -i [\hat H_{S},\rho (t)]+ e^{- i H_{S}t}\dot\rho_{I,S}(t)e^{i H_{S}t}$ the complicated time dependence in the interaction picture disappears and after a straightforward calculation we obtain the following Lindblad equation. Here we reinstate the $\alpha$ index, which sums over all the qubits in the many-body mediator between the masses of our system.
\be
\label{Lindbladian_Schrodinger_many}
\begin{split}
&\dot \rho_{S} (t)= -i [H_{S}+ \sum_{s,\alpha}\Lambda_{s,\alpha}\,,\,\rho_{S} (t)]\\&
+\sum_{s,\alpha} L_{\alpha,s}\rho_{S}(t)L^{\dagger}_{n,s} -\sum_{s,\alpha}\frac{1}{2}\{L^{\dagger}_{\alpha,s}L_{\alpha,s},\rho_{S}(t)\}\,,
\end{split}
\ee
where the Lamb shift and Lindblad jump operators are
\be
\label{Lamb_shift_Lindblad_operators_approx_Schrodinger_2_many}
\begin{split}
\Lambda_{\alpha,+}&=\varepsilon_{+}\,\sigma^{\alpha}_{+}\sigma^{\alpha}_{-}~,~\Lambda_{\alpha,-}=\varepsilon_{-}\,\sigma^{\alpha}_{-}\sigma^{
\alpha
}_{+}\,,\\
L_{\alpha,-}&=\sqrt{\gamma(\omega_{\alpha}(\mb{x}))(n_{B}(\omega_{\alpha}(\mb{x}))+1)}\,\sigma^{\alpha}_{-}\,,\\
L_{\alpha,+}&=\sqrt{\gamma(\omega_\alpha(\mb{x}))n_{B}(\omega_{\alpha}(\mb{x}))}\,\sigma^{\alpha}_{+}\,.
\end{split}
\ee
Here $n_{B}(\omega)= (e^{\omega/T}-1)^{-1}$ represents the Boltzmann thermal occupancy of the bath, and we defined $\epsilon_{s}=s\,{\cal P}\int_{-\infty}^{\infty}d\omega\, J_{s}(\omega)/\omega$, $\gamma(\omega)=(2\pi)D(\omega)g^{2}(\omega)$ and $D(\omega)$ is the bath density of states.

\section{Adiabatic elimination}

\label{app:adiabatic}

The self thermalizing rate of the bath $\Gamma_B$ in the main text is our shortest timescale. Once we have traced out these degrees of freedom, we are left with a Lindblad equation describing the evolution of the massive bodies ($S$) and the many-body mediator ($M$). To proceed to isolate the dynamics of the massive bodies and derive an effective Lindblad equation for just the mass system, we consider the scenario in which the thermalization rate $\gamma(\omega_{\alpha}(\hat x))$ is the fastest timescale when considering dynamics of the  combined SM system. 

In the joint mass-spin system and assuming the masses are slowly moving, the kinetic term of the masses can be neglected.  We then might consider treating {$H_S$} perturbatively after exactly solving the spin master equation. This is a type of adiabatic elimination as used in the quantum optics community~\cite{carmichael2013statistical,carmichael2007statistical,sakurai2020modern,azouit2016adiabatic}. Here we more rigorously show our approximation by appropriate transformation to the Laplace domain and then back. We will ignore the Lamb shift contributions $\Lambda_{\alpha,\pm}$ in Eq.~\eqref{Lamb_shift_Lindblad_operators_approx_Schrodinger_2_many} since they can be canceled by an appropriate shift of $H_{M}$ in \eqref{eq:H-full}.

The general strategy is as follows. Our goal is to solve the Lindblad equation for the mediator qubits, Eq.~\eqref{eq:lindblad}, then insert this solution, leaving a Linblad equation for just the masses. We start with an ansatz for the joint state of the masses and mediator:
\begin{align}
\begin{split}
\label{eq:c-ansatz_app}
\rho_{SM}(t) & = \int d\mb{x} d\mb{x}'  \ket{\mb{x}} \bra{\mb{x}'} \\
& \bigotimes_{\alpha} \left[ c_{\alpha,0}(\mb{x},\mb{x}',t) \ket{0} \bra{0}_{\alpha} + c_{\alpha,1}(\mb{x},\mb{x}',t) \ket{1} \bra{1}_{\alpha} \right].
\end{split}
\end{align}
Here $\mb{x} = \mb{x}_1, \mb{x}_2, \ldots$ represents the coordinates of all the massive bodies. The form of Eq.~\eqref{eq:c-ansatz_app} is preserved by the Lindblad equation, because the Lindblad operators are proportional to position operators. Moreover, any non-zero initial off diagonal elements for each qubit density matrix decay to zero and do not depend on the diagonal elements $\rho_{\alpha,00},\rho_{\alpha,11}$ above, thus the diagonal form of Eq.~\eqref{eq:c-ansatz_app}. 

The normalization condition $\tr \rho_{SM}=1$ implies that the $c_{\alpha}$ coefficients satisfy
\begin{align}
\begin{split}
\label{norm-qubits}
\int d\mb{x} \prod_{\alpha} c_{\alpha,0}(\mb{x},\mb{x},t) + c_{\alpha,1}(\mb{x},\mb{x},t) = 1.
\end{split}
\end{align}
The state $\rho_S = \tr_{M} \rho_{SM}$ of just the masses is
\begin{align}
\label{mass_density_matrix}
\begin{split}
\rho_{S}(t) & = \int d\mb{x} d\mb{x}' \rho_{S}(\mb{x},\mb{x}',t) \ket{\mb{x}} \bra{\mb{x}'}, \\
\rho_{S}(\mb{x},\mb{x}',t) & = \prod_{\alpha} \left[ c_{\alpha,0}(\mb{x},\mb{x}',t) + c_{\alpha,1}(\mb{x},\mb{x}',t) \right].
\end{split}
\end{align}
The time derivative is then given by 
\be
\label{eq:masses_reduced_app}
\dot\rho_{S}(\mb{x},\mb{x}',t)=\sum_{\alpha} C_{\alpha} \rho_{S}(\mb{x},\mb{x}',t),
\ee
where
\be
\label{eq:c_fraction}
 C_{\alpha} = \frac{ \dot{c}_{\alpha,0}(\mb{x},\mb{x}',t) + \dot{c}_{\alpha,1}(\mb{x},\mb{x}',t)}{ c_{\alpha,0}(\mb{x},\mb{x}',t) + c_{\alpha,1}(\mb{x},\mb{x}',t)}.
\ee
Thus, what we want is an equation for this $c$ fraction. What we will do is compute the $c$ fraction in the ``late-time'' limit $t \gtrsim 1/\gamma$, and then use the result to reduce Eq.~\eqref{eq:masses_reduced_app} into the Lindblad equation for the masses, as in Eq.~\eqref{eq:lindblad-final}.

Inserting Eq. \eqref{eq:c-ansatz_app} into the Lindblad equation \eqref{eq:lindblad} one finds a differential equation for the $c_{\alpha}$ coefficients. These equations factor over the label $\alpha$ so from here out we suppress the $\alpha$ subscripts. The equations are
\be
\label{eq:dotc}
\dot{\mb{c}}  = \Lambda \mb{c}, \ \ \ \Lambda = \Lambda_0 + \delta \Lambda, \ \ \ \mb{c} = \begin{pmatrix} c_0 \\ c_1 \end{pmatrix}.
\ee
We emphasize that everything appearing here is a function of both the ket and bra variables $\mb{x},\mb{x}'$. The specific form of the matrix $\Lambda$ will differ in the local and non-local models. However, in both cases, we decompose the dynamics into the part $\Lambda_0$ which drives the qubits to a thermal state with rate $\gamma$, and a part $\delta \Lambda$ which generates perturbations around the thermal state. This is suggestive of the kind of ``perturbation theory" we are about to engage in. The $\Lambda_0$ matrix will be considered as the dominant term, while $\delta \Lambda$ will be treated as a a kind of perturbation. 

Our goal is to find the $c$ ratio in Eq.~\eqref{eq:masses_reduced_app} in the limit of times long compared to the decay rate $t \gtrsim \gamma^{-1}$. We will first solve the problem ignoring $\delta \Lambda$ and then include its effects. Let
\be
\Lambda_0 \mb{v}_s = \lambda_s \mb{v}_s, \ \ \ \Lambda_d \mb{v}_d = \lambda_d \mb{v}_d
\ee
denote the eigenvectors and eigenvalues of $\Lambda_0$. The notation will refer to a  decaying mode with $\lambda_d \sim -\gamma$ and a quasi-steady state mode with $\lambda_s \sim \gamma^0$. The existence of this decaying and steady state mode structure is not guaranteed in a general Lindbladian system, but it exists in the models in this paper, as we will see explicitly below. Ignoring $\delta \Lambda$, the general solution to $\dot{\mb{c}} = \Lambda_0 \mb{c}$ is then a linear combination of $e^{\lambda_s t}$ and $e^{\lambda_d t}$, so in particular the decaying mode is exponentially damped. 

Now we include the effects of the $\delta \Lambda$ term. Let $R = (\mb{v}_s,\mb{v}_d)$ be the matrix that diagonalizes $\Lambda_0$. We rotate everything into this basis as usual
\be
\tilde{\Lambda} = R^{-1} \Lambda R, \ \ \ \tilde{\mb{c}} = R^{-1} \mb{c} = \begin{pmatrix} \tilde{c}_s \\ \tilde{c}_d \end{pmatrix},
\ee
and use $s,d$ to refer to the components in this basis. The equation of motion for the state components becomes $d\tilde{\mb{c}}/dt = \tilde{\Lambda} \tilde{\mb{c}}$, and $\tilde{\Lambda} = \tilde{\Lambda}_0 + \delta \tilde{\Lambda}$, with
\be
\tilde{\Lambda}_0 = \begin{pmatrix} \lambda_s & 0 \\ 0 & \lambda_d \end{pmatrix}, \ \ \ \delta \tilde{\Lambda} = \begin{pmatrix} \delta \tilde{\Lambda}_{ss} & \delta \tilde{\Lambda}_{sd} \\ \delta \tilde{\Lambda}_{ds} & \delta \tilde{\Lambda}_{dd} \end{pmatrix}.
\ee
The essential observation of the adiabatic elimination procedure comes from analyzing the equation for the decaying component,
\be
\frac{d \tilde{c}_d}{dt} = \left[ \lambda_d + \delta \tilde{\Lambda}_{dd} \right] \tilde{c}_d + \delta \tilde{\Lambda}_{ds} \tilde{c}_s
\ee
in the limit of $t \gtrsim 1/\lambda_d \sim 1/\gamma$. In this limit, the derivative on the left-hand-side goes to zero, and we can just write down the algebraic solution
\be
\label{eq:cd_late}
\tilde{c}_d \to Z \tilde{c}_{s}, \ \ \ Z = - \frac{\delta \tilde{\Lambda}_{ds}}{\lambda_d + \delta \tilde{\Lambda}_{dd}}.
\ee
This can be justified more systematically by analyzing the equations in the Laplace domain. With this result, we then go back to the equation for the quasi-steady state mode, and find
\be
\label{eq:cs_soln}
\frac{d \tilde{c}_s}{dt} = \left[ \lambda_s + \delta \tilde{\Lambda}_{ss} + \delta \tilde{\Lambda}_{sd} Z \right] \tilde{c}_s.
\ee
This, finally, is enough to get our desired prefactor in Eq.~\eqref{eq:masses_reduced_app}: from Eq.~\eqref{eq:cd_late} and the definition of $\tilde{\mb{c}}$, we have
\begin{align}
\begin{split}
c_0 + c_1 & = \left[ R_{0s} + R_{1s} \right] c_s + \left[ R_{0d} + R_{1d} \right] c_d \\
& = \left[ R_{0s} + R_{1s} + Z \left( R_{0d} + R_{1d} \right) \right] c_s
\end{split}
\end{align}
and then using Eq.~\eqref{eq:cs_soln}, we obtain
\be
\label{eq:cfrac}
C = \frac{\dot{c}_0 + \dot{c}_1}{c_0 + c_1} = \left[ \lambda_s + \delta \tilde{\Lambda}_{ss} - \frac{\delta \tilde{\Lambda}_{sd} \delta \tilde{\Lambda}_{ds}}{\lambda_d + \delta \tilde{\Lambda}_{dd}} \right].
\ee
This expression works for all $\alpha$ and $\mb{x},\mb{x}'$, and gives the prefactor in Eq.~\eqref{eq:masses_reduced_app}. 

Finally, we need to convert the result into a Lindblad equation of the form of Eq.~\eqref{eq:lindblad-final}, which means identifying the correct effective Hamiltonian and jump operators from the position-space expression in Eq.~\eqref{eq:masses_reduced_app}. This turns out to be straightforward since any appearance of $\mb{x}$ has to correspond to a left-acting operator, while any appearance of $\mb{x}'$ is a right-acting operator. This will be easier to understand with more explicit expressions in the examples below.

\subsection{Local model} 
It turns out that following the adiabatic elimination procedure is a bit more straightforward in the local model, because things can be expanded around large values of the chemical potential $\mu$, so we begin with that case.

Inserting the ansatz of Eq.~\eqref{eq:c-ansatz_app} into the system-mediator Lindbladian in Eq.~\eqref{eq:lindblad}, we find the matrices $\Lambda_0$ and $\delta \Lambda$,
\begin{equation}
\Lambda_0 = \gamma_{\rm th} \begin{pmatrix} 
 -e^{\mu/T}&  1 \\
e^{\mu/T}  &  - 1 \end{pmatrix}
\end{equation}
and
\begin{equation}
    \delta\Lambda = \left( \begin{array}{cc}
 \frac{\gamma_{\rm th} e^{\mu/T}}{2}(2-e^{- \omega/T}-e^{-\omega^{\prime}/T})& 0 \\
\gamma_{\rm th} e^{ \mu/T}(e^{-(\omega+\omega^{\prime})/2T}-1) & - i(\omega - \omega^{\prime})
\end{array} \right),
\end{equation}
where we are using the shorthand notation $\omega = \omega_{\alpha}(\mb{x})$, $\omega' = \omega_{\alpha}(\mb{x}')$. The matrix $\Lambda_0$ has eigenvalues 
\be
\lambda_s = 0, \ \ \ \lambda_d = -\gamma_{\rm th} (1+e^{\mu/T}),
\ee
and corresponding eigenvectors
\be
\mb{v}_s = \begin{pmatrix} e^{-\mu/T} \\ 1 \end{pmatrix}, \ \ \ \mb{v}_d = \begin{pmatrix} -1 \\ 1 \end{pmatrix}.
\ee
Using these vectors we can construct the $R$ matrix that rotates into this basis, 
\begin{equation}
R = \begin{pmatrix}
 e^{-\mu/T} & -1\\
1 & 1 \end{pmatrix}
\end{equation}
From this, we can find the elements of the perturbation matrix $\delta \Lambda$ in the $s,d$ basis. They are
\begin{align}
\begin{split}
 \delta \tilde{\Lambda}_{dd} &= - \gamma_{\rm th} W_1 + \gamma_{\rm th} W_2 - i (1-\sigma_{*})(\omega - \omega^{\prime}) \\
\delta \tilde{\Lambda}_{ds} &= -\frac{(1-\sigma_{*})}{\sigma_{*}}\left(\delta \tilde{\Lambda}_{dd} + i (\omega - \omega^{\prime})\right) \\
\delta \tilde{\Lambda}_{ss} & = - \gamma_{\rm th} W_3 - i  \sigma_{*}(\omega - \omega^{\prime}) \\
\delta \tilde{\Lambda}_{sd} &= -\frac{\sigma_{*}}{(1-\sigma_{*})}\left(\delta \tilde{\Lambda}_{ss} + i (\omega - \omega^{\prime})\right)
\end{split}
\end{align}
in terms of the coefficients
\begin{align}
\begin{split}
W_1 &= \sigma_{*} \left( e^{- (\omega + \omega^{\prime})/2T} - e^{\mu/T} \right) \\
W_2 &= \frac{\sigma_{*}}{2} \left( 2 - e^{(\mu - \omega)/T} - e^{(\mu - \omega^{\prime})/T} \right) \\
W_3 &= \frac{\sigma_{*}}{2} \left( e^{-\omega/2T} - e^{-\omega^{\prime}/2T} \right)^2.
\end{split}
\end{align}
These expressions are somewhat unwieldy, but they are enough to determine the $\dot{\rho}_S$ evolution, using Eq.~\eqref{eq:cfrac} in Eq.~\eqref{eq:masses_reduced_app}, i.e., 
\begin{align}
C = \frac{\dot{c}_0 + \dot{c}_1}{c_0 + c_1}  = \left[ \lambda_s + \delta \tilde{\Lambda}_{ss} - \frac{\delta \tilde{\Lambda}_{sd} \delta \tilde{\Lambda}_{ds}}{\lambda_d + \delta \tilde{\Lambda}_{dd}} \right]. 
\end{align}
We now expand this around $|\omega|, |\omega'| \ll \mu, T$, to quadratic order, and find $C = C_+ + C_0 + C_-$, where
\begin{align}
\begin{split}
\label{eq:Cs-local}
C_0 & = -i\left(\sigma_* \delta \omega -\frac{\sigma_{*}(1-\sigma_{*})}{2T} (\omega^2 - \omega'^{2}) \right)\sim \gamma_{\rm th}^{0} \\
C_+ & =  -\frac{\sigma_*}{8 T^2}  \gamma_{\rm th} \delta\omega^2 \sim \gamma_{\rm th} \\
C_- & =  -\frac{1}{\gamma_{\rm th}} \sigma_* (\sigma_* - 1)^2 \delta \omega^2 \sim \gamma_{\rm th}^{-1},
\end{split}
\end{align}
with $\delta \omega = \omega - \omega'$. Now we want to match these terms to a Lindblad equation, i.e., identify the effective Hamiltonian and noise operators in Eq.~\eqref{eq:lindblad-final}.

First we analyze the $C_0$ terms, which we will see are the $\omega$-space representation of an effective Hamiltonian that gives rise to a coherent Newtonian gravitational interaction. Consider the first term. Summing over the qubits $\alpha$, this contributes to the position-space density matrix evolution as
\be
\dot{\rho}_S(\mb{x},\mb{x}') = -i \sigma_* \sum_{\alpha} \left( (\omega_{\alpha}(\mb{x}) - \omega_{\alpha}(\mb{x}') \right) \rho_S(\mb{x},\mb{x}').
\ee
Now compare this to a Hamiltonian operator $H = \sigma_* \sum_{\alpha} \omega(\mb{x})$, where here $\mb{x}$ is an operator. Such a Hamiltonian would similarly give a contribution in position space
\begin{align}
\begin{split}
\dot{\rho}_S(\mb{x},\mb{x}') & = -i \braket{\mb{x} | [H,\rho_S] | \mb{x}'} \\
& = -i \sum_{\alpha} \left( (\omega_{\alpha}(\mb{x}) - \omega_{\alpha}(\mb{x}') \right) \rho_S(\mb{x},\mb{x}').
\end{split}
\end{align}
Thus, we see that the first term in our adiabatic evolution result Eq.~\eqref{eq:Cs-local} is exactly the same as an effective Hamtiltonian of this form. This worked because a position operator $\mb{x}$ acts in position space as $\mb{x}$ on the left and as $\mb{x}'$ on the right. The same logic works for the second term in $C_0$ in Eq.~\eqref{eq:Cs-local}. We thus conclude that the contribution from $C_0$ in Eq.~\eqref{eq:masses_reduced_app} is equivalent to an effective Hamiltonian
\be
V_{N,{\rm ent}} = \sigma_* \sum_{\alpha} \omega_{\alpha}(\mb{x}) - \frac{\sigma_*(1-\sigma_*)}{2T} \sum_{\alpha} \omega_{\alpha}^2(\mb{x}),
\ee
which is exactly what was claimed in Eq.~\eqref{eq:VN_local}. In the continuum limit and with the identification of Eq.~\eqref{eq:G-sigma}, this becomes exactly a constant plus the Newtonian potential operator, as shown in the main text.

Similar logic applies to the $C_{\pm}$ terms, except that these take the structure of Lindblad noise operators, not Hamiltonian operators. Consider the $C_+$ terms first. Again summing over all the qubits, in position space these give a contribution to the density matrix evolution
\be
\dot{\rho}_S(\mb{x},\mb{x}') = - \frac{\sigma_*}{8 T^2} \gamma_{\rm th} \sum_{\alpha} (\omega_{\alpha}(\mb{x}) - \omega_{\alpha}(\mb{x}'))^2 \rho_S(\mb{x},\mb{x}').
\ee
Now we compare this to the effect of a (Hermitian) Lindblad operator of the form $K_{\alpha,+} = K_{\alpha,+}(\omega_{\alpha}(\mb{x}))$, which would contribute
\begin{align}
\begin{split}
& \dot{\rho}_S(\mb{x},\mb{x}') = \sum_{\alpha} \braket{\mb{x} | K_{\alpha,+} \rho_S K_{\alpha,+} - \frac{1}{2} \left\{ K_{\alpha,+}^2, \rho_S \right\} | \mb{x}'} \\
& -\frac{1}{2} \sum_{\alpha} \big[  K_{\alpha,+}(\omega_{\alpha}(\mb{x})) -  K_{\alpha,+}(\omega_{\alpha}(\mb{x}')) \big]^2 \rho_S(\mb{x},\mb{x}').
\end{split}
\end{align}
Thus we see that the $C_+$ terms arising in Eq.~\eqref{eq:masses_reduced_app} are equivalent to Lindblad operators of the form
\be
K_{\alpha,+}(\omega_{\alpha}(\mb{x})) = \sqrt{\frac{\sigma_*}{4 T^2}} \omega_{\alpha}(\mb{x}).
\ee
The exact same logic works for the $C_-$ terms, and leads to the identification of the Lindblad operators
\be
K_{\alpha,-}(\omega_{\alpha}(\mb{x})) = \sqrt{ \frac{2 \sigma_* (\sigma_* - 1)^2}{\gamma_{\rm th}}} \omega_{\alpha}(\mb{x}).
\ee
This is precisely as advertised in Eq.~\eqref{eq:lindblad-op-local}.

To summarize, we have a conservative term, i.e., an effective Hamiltonian
\be
H_{\rm eff} = \sum_{\alpha} H_{\alpha} = \sigma_* \sum_{\alpha} \omega_{\alpha} - \frac{\sigma_* ( 1- \sigma_*)}{2T} \sum_{\alpha}  \omega_{\alpha}^2,
\ee 
which is exactly what we expected from the free energy (as opposed to just the internal energy) in the classical expression given in Eq.~\eqref{eq:A-local-taylor}. We also have a damping term of Lindblad form that looks much like position-based dephasing mechanisms due to the localization of the masses by the $\omega_{\alpha}(\mb{x})$ operators. As discussed extensively in the main text, the dephasing component has two terms: one of which goes with $\gamma_{\rm th}$ and the other going as $1/\gamma_{\rm th}$. The first is due to the masses being measured by the qubits, while the second is due to the fluctuating force on the masses due to the spins in the background flipping with a correlation time $1/\gamma_{\rm th}$.

\subsection{Non-local model}

In the non-local model, we have to make a slightly different set of approximations. Again inserting the anstaz in Eq.~\eqref{eq:c-ansatz_app} into the Lindblad equation in Eq.~\eqref{eq:lindblad}, we find
\begin{align}
\begin{split}
(\Lambda_0)_{11}&=-\frac{1}{2}\left[\gamma (\omega)(n_{B}(\omega)+1) + \gamma(\omega')(n_{B}(\omega')+1)\right],\\
(\Lambda_0)_{10}&=\sqrt{\gamma(\omega) n_{B}(\omega)}\sqrt{\gamma(\omega') n_{B}(\omega'}),\\
(\Lambda_0)_{01}&=\sqrt{\gamma(\omega)[n_B(\omega)+1]}\sqrt{\gamma(\omega')[n_B(\omega')+1]},\\
(\Lambda_0)_{00}&=- \frac{1}{2} \left[ \gamma(\omega) n_B(\omega) + \gamma(\omega') n_B(\omega') \right],
\end{split}
\end{align}
and
\be
\delta \Lambda = \begin{pmatrix} 0 & 0 \\ 0 & -i (\omega - \omega') \end{pmatrix}.
\ee
Here we again use the shorthand $\omega=\omega(\mb{x})$, $\omega'=\omega(\mb{x}')$.

The unperturbed $\Lambda_0$ is easily diagonalized. Unlike the local model, there is not an exact zero eigenvalue, i.e., there is not an exact steady state even in the absence of the $\delta \Lambda$ term. The eigenvalues are 
\begin{align}
\begin{split}
\label{eq:lambdas-nl}
\lambda_{s} & = -\frac{1}{4}\bigg[\gamma(\omega)(2 n_B(\omega)+1)+ \gamma(\omega')(2 n_B(\omega')+1) \\
& - \alpha (\omega,\omega')\bigg] \approx 0 \\
\lambda_{d} & =-\frac{1}{4}\bigg[\gamma(\omega)(2 n_B(\omega)+1)+ \gamma(\omega')(2 n_B(\omega')+1) \\
& + \alpha (\omega,\omega')\bigg] \approx- \gamma(\omega) \left[ 1 + 2 n_B(\omega) \right].
\end{split}
\end{align}
Here the approximations show the behavior as $\omega' \to \omega$. The idea is that for small superpositions ($\mb{x}' \approx \mb{x}$, thus $\omega' \approx \omega$), we see that $\lambda_d \approx -\gamma$ and $\lambda_s \approx 0$, in line with the general structure discussed above. The factor
\begin{align}
\begin{split}
\label{alpha_formula}
\alpha(\omega,\omega') &=  \sqrt{(\gamma(\omega)+\gamma(\omega'))^2 + 16\gamma(\omega)\gamma(\omega')\beta(\omega,\omega')}\\
& \approx 2\gamma(\omega)\left(1 + 2 n_B(\omega)\right),
\end{split}
\end{align}
with
\be
\beta(\omega,\omega')=\sqrt{n_{B}(\omega)(n_{B}(\omega)+1)n_{B}(\omega')(n_{B}(\omega')+1)}
\ee
will appear repeatedly. The corresponding eigenvectors are
\begin{align}
\begin{split}
\mb{v}_{s} & =  \begin{pmatrix} \frac{\gamma(\omega)+\gamma(\omega') + \alpha(\omega,\omega')}{4 \sqrt{[\gamma(\omega)n_{B}(\omega )] [\gamma(\omega') n_{B}(\omega')]}} \\ 1 \end{pmatrix} \approx \begin{pmatrix} \frac{n_{B}(\omega)+1}{n_{B}(\omega)} \\ 1 \end{pmatrix} \\
\mb{v}_{d} & =  \begin{pmatrix} \frac{\gamma(\omega)+\gamma(\omega') - \alpha(\omega,\omega')}{4 \sqrt{[\gamma(\omega)n_{B}(\omega )] [\gamma(\omega') n_{B}(\omega')]}} \\ 1 \end{pmatrix} \approx \begin{pmatrix} -1 \\ 1 \end{pmatrix}.
\end{split}
\end{align}
While the lack of a zero eigenvalue means that there is not a strict steady state unless $\omega = \omega'$, the $\mb{v}_s$ state still decays strictly slower than the decaying eigenvector $\mb{v}_d$, and in this sense provides a quasi-steady state for times $t \gtrsim 1/\gamma$. 

To get to the Lindblad equation for the masses, we again use Eq.~\eqref{eq:cfrac} in Eq.~\eqref{eq:masses_reduced_app}. Moving to the $s,d$ basis, we have
\begin{align}
\label{eq:tilde_lambda_nl}
\delta \tilde{\Lambda} =   \frac{i}{2} \delta \omega \begin{pmatrix} \frac{\gamma(\omega)+\gamma(\omega') - \alpha(\omega,\omega')}{\alpha(\omega,\omega')} & \frac{\gamma(\omega)+\gamma(\omega') - \alpha(\omega,\omega')}{\alpha(\omega,\omega')} \\ \frac{-\gamma(\omega)- \gamma(\omega') - \alpha(\omega,\omega')}{\alpha(\omega,\omega')} & \frac{-\gamma(\omega)-\gamma(\omega') - \alpha(\omega,\omega')}{\alpha(\omega,\omega')} \end{pmatrix},
\end{align}
where $\delta \omega = \omega - \omega'$. For small superpositions $\delta \omega \approx 0$, one can expand the $c$ fraction in Eqs.~\eqref{eq:c_fraction},~\eqref{eq:cfrac} in terms of the individual terms, which scale like
\begin{align}
\begin{split}
\label{eq:c_frac_nl}
C_{0} & = \delta \tilde{\Lambda}_{ss} \sim \gamma^0 \\
C_{+} & = \lambda_s \sim \gamma \\
C_{-} & = \frac{\delta \tilde{\Lambda}_{sd} \delta \tilde{\Lambda}_{ds}}{\lambda_d + \delta \tilde{\Lambda}_{dd}} \sim \gamma^{-1}.
\end{split}
\end{align}
Let us examine each of these terms and map them into a Lindblad structure.

First, consider the term $C_{0} = \delta \tilde{\Lambda}_{ss}$, which we will see leads to the effective Newtonian interaction. The explicit expression is
\be
\label{eq:C0-nl}
C_0 = -i \frac{ n_{B}(\omega)}{2 n_{B}(\omega )+1} \delta \omega + i\frac{n_{B}'(\omega )}{2 (2 n_{B}(\omega )+1)^2} \delta \omega^2 + \mathcal{O}(\delta \omega^3).
\ee
We need to evaluate this summed over all the qubits $\alpha$, as in Eq.~\eqref{eq:masses_reduced_app}. The first term in Eq.~\eqref{eq:C0-nl} gives a contribution
\begin{align}
\begin{split}
\label{eq:C0_nl_first}
& -i \sum_{\alpha}  \frac{ n_{B}(\omega_{\alpha})}{2 n_{B}(\omega_{\alpha})+1} \delta \omega_{\alpha} =  -i \sum_{\alpha} n_{FD}(\omega_{\alpha}) \delta \omega_{\alpha} \\
& \ \ \ \ \ \  \approx -i \frac{\nabla f(\mb{x}) \cdot \delta \mb{x}}{f^2(\mb{x})} \int d\omega \,  n_{FD}(\omega) \omega \\
& \ \ \ \ \ \ = -i \frac{\pi^2 T^2 \ell^2}{12} \frac{\hat{\mb{x}} \cdot \delta \mb{x}}{|\mb{x}|^2}.
\end{split}
\end{align}
To get the first line, we used $n_{FD} = n_B/(2n_B + 1)$. To get the second line, we used $\omega_{\alpha}(\mb{x}) = \alpha f(\mb{x})$ to write $\delta \omega_{\alpha} = \omega_{\alpha}(\mb{x}) - \omega_{\alpha}(\mb{x}') \approx \omega_{\alpha} \nabla f(\mb{x}) \cdot \delta \mb{x}/f(\mb{x})$, as in the main text when estimating noise effects, and took the continuum limit. The third line is the simple result of the integral. Now let's compare this term in $\dot{\rho}_S$ to that coming from a Newtonian potential interaction $V_N = -G_N m_1 m_2/|\mb{x}|$. This would give
\begin{align}
\begin{split}
\label{eq:C0_nl_ham}
& \dot{\rho}_S(\mb{x},\mb{x}')  = -i \braket{\mb{x} | [V_{N},\rho_{S}] | \mb{x}'} \\
& = -i\rho_{S}(\mb{x},\mb{x}',t) \left[-\frac{G_{N}m_{1}m_{2}}{|\mb{x}|}+ \frac{G_{N}m_{1}m_{2}}{|\mb{x}'|} \right] \\
& = -i \rho_{S}(\mb{x},\mb{x}',t) \frac{G_N m_1 m_2}{|\mb{x}|} \\
& \times \left[ - \frac{\hat{\mb{x}} \cdot \delta \mb{x}}{|\mb{x}|} + \frac{1}{2} \left( \frac{3 (\hat{\mb{x}} \cdot \delta \mb{x})^2}{|\mb{x}|^2} - \frac{\delta \mb{x}^2}{|\mb{x}|^2} \right) + \cdots \right].
\end{split}
\end{align}
In the last line, we again made use of the assumption that we are looking at a small superposition, in the sense that $\mb{x}' = \mb{x} + \delta \mb{x}$, with $|\delta \mb{x}| \ll |\mb{x}|$. Now, invoking the usual identification $\pi^2 T^2 \ell^2/12 = G_N m_1 m_2$ [Eq.~\eqref{eq:T-G}], we see that the first term here precisely matches the one computed from the microscopic adiabatic elimination, Eq.~\eqref{eq:C0_nl_first}. To get the quadrupole term in Eq.~\eqref{eq:C0_nl_ham}, one needs the second-order derivative term 
\be
\delta \omega = \partial_i \omega(\mb{x}) \delta x^i + \frac{1}{2} \partial_i \partial_j \omega(\mb{x}) \delta x^i \delta x^j + \cdots
\ee
coming from the first term in Eq.~\eqref{eq:C0-nl}, as well as the term $\sim \delta \omega^2 \sim (\partial_i \omega(\mb{x}) \delta x^i)^2$ from the second term in Eq.~\eqref{eq:C0-nl}. Similar manipulations then show that the $\dot{\rho}_S$ contribution from these second order terms in Eq.~\eqref{eq:C0-nl} also exactly match the quadrupole term in Eq.~\eqref{eq:C0_nl_ham}. We thus conclude that, at least to this order in small superposition size, the entropic model has an effective Hamiltonian $H_{\rm eff}$ containing precisely the usual coherent, Newtonian gravitational interaction. It would interesting to to extend this to higher orders.

We can follow the same logic to identify Lindblad noise operators coming from the $C_{\pm}$ contributions. First consider the term $C_+ = \lambda_s$. This leads to a damping contribution of $\mathcal{O}(\gamma)$. The explicit expression is obtained by expanding Eq.~\eqref{eq:lambdas-nl} to order $\delta \omega^2$. This gives
\begin{align} 
\begin{split}
& C_+  = -\frac{1}{8} \Bigg[2 \gamma'(\omega ) n_B'(\omega ) + \frac{2 n_B(\omega ) (n_B(\omega )+1) \gamma '(\omega )^2}{\gamma (\omega ) (2 n_B(\omega )+1)} \\
& + \frac{\gamma (\omega ) (2 n_B(\omega ) (n_B(\omega )+1)+1) n_B'(\omega )^2}{n_B(\omega ) (n_B(\omega )+1) (2 n_B(\omega )+1)}  \Bigg] \delta \omega^2  + \mathcal{O}(\delta \omega^3).
\end{split} 
\end{align}
To make progress at this stage, we have to assume a specific bath spectral density. As discussed in the main text, we take this to be Ohmic $\gamma(\omega)=\zeta \omega$, with $\zeta$ a dimensionless constant. This gives, for each qubit $\alpha$, a contribution to the $\dot{\rho}_S$ equation of motion of the form
\be
\label{eq:Cp_nl_direct}
C_{\alpha,+} = - \frac{\zeta}{T} g_+\left( \frac{\omega_{\alpha}}{T} \right) \delta \omega_{\alpha}^2,
\ee
where the dimensionless function $g_{+}$ is
\begin{align}
\begin{split}
\label{eq:gplus}
g_{+}(\nu) &= \frac{1}{64 \nu} \text{csch}^{3}(\nu/2) \text{sech}(\nu/2) \\
& \times \left[-2 + (2+ \nu^2)\cosh(\nu) -2\nu\, \text{sinh}(\nu)\right].
\end{split}
\end{align}
Again this needs to be summed over the qubits $\alpha$ to obtain the contribution to the $\dot{\rho}_S$ equation. Now, we want to compare this to the effect of some Lindblad operators of the form $K_{\alpha,+} = K_{\alpha,+}(\omega_{\alpha}(\mb{x}))$. These contribute to the position-space $\dot{\rho}_S$ equation in the form
\begin{align}
\begin{split}
\label{eq:Cp_nl_lindblad}
& \dot{\rho}_S(\mb{x},\mb{x}')  \\
& = -\frac{1}{2} \sum_{\alpha} \left[ K_{\alpha,+}(\omega_{\alpha}(\mb{x})) - K_{\alpha,+}(\omega_{\alpha}(\mb{x}')) \right]^2 \rho_S(\mb{x},\mb{x}') \\
& \approx -\frac{1}{2} \sum_{\alpha} \left[ K'_{\alpha,+}(\omega_{\alpha}(\mb{x})) \right]^2 \delta \omega_{\alpha}^2 \dot{\rho}_S(\mb{x},\mb{x}').
\end{split}
\end{align}
In the first line, we assumed that the $K$ operators are Hermitian for simplicity, and in the second line used our assumption of a small superposition to expand in derivatives $' = \partial_{\omega}$. Comparing Eqs.~\eqref{eq:Cp_nl_direct} and \eqref{eq:Cp_nl_lindblad}, we identify the effective Lindblad operators as
\be
K_{\alpha,+}(\omega_{\alpha}(\mb{x})) = \sqrt{2 T \zeta}  \int_0^{\omega_{\alpha}(\mb{x})/T} d\nu \, g_+(\nu),
\ee
as shown in Eq.~\eqref{eq:Kpm}.

Finally, we analyze the third term $C_-$ in Eq.~\eqref{eq:c_frac_nl}, which gives a damping effect at $\mathcal{O}(\gamma^{-1})$. The explicit expression is
\be
\label{eq:tilde-lambda-sd_nl}
C_- = \frac{\delta \tilde{\Lambda}_{sd} \delta \tilde{\Lambda}_{ds}}{\lambda_d + \delta \tilde{\Lambda}_{dd}}  = \frac{n(\omega ) (n(\omega )+1)}{\gamma (\omega ) (2 n(\omega )+1)^3} \delta \omega^2 + \mathcal{O}(\delta \omega^3).
\ee
The logic is then identical to the previous paragraph. Invoking the Ohmic form of $\gamma(\omega) = \zeta \omega$ and matching to a Lindblad operator $K_{\alpha,-} = K_{\alpha,-}(\omega_{\alpha}(\mb{x}))$, we find the Lindblad operator
\be
K_{\alpha,-}(\omega_{\alpha}(\mb{x})) = \sqrt{\frac{2 T}{\zeta}} \int_0^{\omega_{\alpha}(\mb{x})/T} d\nu \, g_-(\nu),
\ee
where the dimensionless function $g_-$ is
\be
\label{eq:gminus}
g_{-}(\nu)=\frac{2}{\nu}\text{csch}^{3}(\nu)\text{sinh}^{4}(\nu/2),
\ee
again as shown in Eq.~\eqref{eq:Kpm} in the main text.

\section{Entanglement witness in the local model}
\label{app:witness}
Here we record the exact expressions for both the gravitational phases and decoherence rates that enter the entanglement witness calculation in the local model in Sec.~\ref{sec:entanglement}. The complications here arise from the fact that experiments can plausibly probe regimes where superpositions $\delta x/a \gtrsim 1$ and $\delta x \approx d$. In the former case, in particular, the local model predicts small deviations from an exact $1/r$ potential, as in Eq.~\eqref{eq:Omega}.

From the form of the Lindblad operators in Eq. \eqref{eq:lindblad-op-local} we obtain 
\bes{
\label{eq:local_dm_elements}
\rho_{LL,RL}&= \frac{e^{-i \phi_{RL}}}{4}e^{-\Gamma_{0} \Delta t},~\rho_{LR,RR}=\frac{e^{i \phi_{LR}}}{4}e^{- \Gamma_{0} \Delta t},\\
\rho_{LR,RL}&=\frac{e^{i(\phi_{RL}-\phi_{LR})}}{4} e^{-\Gamma_{+} \Delta t},~\rho_{LL,RR}=\frac{1}{4}e^{-\Gamma_{-} t},
}
where the phases are
\bes{
\label{eq:phases-local-exact}
\phi_{RL}&= G_{N}m^{2}\left(\frac{s(d-\delta x)}{d-\delta x}- \frac{s(d)}{d}\right),\\
\phi_{LR}&= G_{N}m^{2}\left(\frac{s(d+\delta x)}{d+\delta x}- \frac{s(d)}{d}\right),
}
and the decay rates are 
\bes{
\label{eq:decay-local-exact}
\Gamma_{0}&= \frac{\pi^2 \kappa}{a^4}m^2L^4 \left(1- \pi a \frac{s(\delta x)}{\delta x}\right),\\
\Gamma_{\pm}&=  \frac{2\pi^2 \kappa}{a^4}L^4 m^{2}\times \bigg[\big(1- \pi a \frac{s(\delta x)}{\delta x}\big) \\& \pm\frac{\pi a}{2}  \left(\frac{s(d+\delta x)}{d+ \delta x}+ \frac{s(d-\delta x)}{d-\delta x} - 2\frac{s(d)}{d}\right)\bigg].
}
The function $s(x)=1 - (2/\pi)\arctan(2 a/x)$ was defined in Eq. \eqref{eq:f_local_dfn} and $\kappa\equiv \frac{\sigma_{*}\gamma_{\rm th}}{4T^2}+ \frac{2}{\gamma_{\rm th}}\sigma_{*}(\sigma_{*}-1)^{2}$.
The minimal decoherence, shown in Fig.~\ref{fig:W}, happens at $\kappa_{\rm min}=\sqrt{2}G_{N}a^{3}/(\pi^3 L^{4})$ where we used Eq. \eqref{eq:G-sigma}. From Eq.~\eqref{eq:local_dm_elements} and Eq.~\eqref{eq:Wt} we finally obtain
\bes{
\label{W_entropic_local_app}
\left| \braket{W} \right|= \frac{1}{2}\big|& e^{-\Gamma_{+} \Delta t}\cos(\phi_{LR}-\phi_{RL}) -e^{-\Gamma_{-} \Delta t} \\& +e^{- \Gamma_{0} \Delta t}\left(\cos\phi_{RL} - \cos \phi_{LR}\right)\big|, 
}
which is the formula shown in Eq.~\eqref{W_entropic_local}.

\end{document}